\documentclass[12pt]{article}
\usepackage{amssymb,slashed,graphicx,color,epsf}

\def\nn{\nonumber}

\def\be{\begin{equation}}
\def\ee{\end{equation}}
\def\ben{\begin{displaymath}}
\def\een{\end{displaymath}}
\def\bea{\begin{eqnarray}}
\def\eea{\end{eqnarray}}

\def\ft#1#2{{\textstyle {\frac{#1}{#2}} }}

\makeatletter \@addtoreset{equation}{section} \makeatother

\def\a{\alpha}

\newcommand{\w}[1]{\\[0.#1cm]}

\def\eq#1{(\ref{#1})}
\def\ft#1#2{{\textstyle{{\scriptstyle #1}\over {\scriptstyle #2}}}}
\newcommand{\hoch}[1]{$\, ^{#1}$}
\def\fft#1#2{{\frac{#1}{#2}}}
\def\tr{{\rm tr} }
\def\kur{\rho}
\def\nk{{\cal O}_{nk}}

\def\beps{{\bar\epsilon}}

\newcommand{\tamphys}{\it George and Cynthia Woods Mitchell  Institute
for Fundamental Physics and Astronomy,\\
Texas A\&M University, College Station, TX 77843, USA}

\newcommand{\auth}{
R. Percacci\hoch{\sharp},
M.J. Perry\hoch{\ddagger},
C.N. Pope\hoch{\dagger,\ddagger}
and E. Sezgin\hoch{\dagger}
}

\begin{document}

\begin{flushright}
\hfill{
MIFPA-13-04}\\
\hfill{DAMTP 2013-6}

\end{flushright}

\vspace{25pt}

\begin{center}

{\large {\bf Beta Functions of Topologically Massive Supergravity}}

\vspace{25pt}
\auth

\vspace{10pt}
\hoch{\sharp}{\it  SISSA, via Bonomea 265, Trieste, Italy and INFN, Sezione di Trieste.}

\vspace{10pt}
\hoch{\dagger}{\tamphys}

\vspace{10pt}
\hoch{\ddagger}{\it  DAMTP, Centre for Mathematical Sciences,
 Cambridge University,\\  Wilberforce Road, Cambridge CB3 0WA, UK}

\vspace{40pt}

\underline{ABSTRACT}
\end{center}

We compute the one-loop beta functions of the cosmological constant, Newton's constant and the topological mass in topologically massive supergravity in three dimensions. We use a variant of the proper time method supplemented by a simple choice of cutoff function. We find that  the dimensionless coefficient of the Chern-Simons term, $\nu$, has vanishing beta function. The flow of the cosmological constant and Newton's constant depends on $\nu$; we study analytically the structure of the flow and its fixed points in the limits of small and large $\nu$.

\vspace{15pt}

\thispagestyle{empty}

\pagebreak


\tableofcontents

\newpage


\section{Introduction}


Topologically massive gravity (TMG) \cite{Deser:1981wh} is described by a Lagrangian in three dimensions
consisting of the Einstein-Hilbert term,  cosmological term and Lorentz Chern-Simons term. Positivity of the
energy for the black hole solution requires that Newton's constant $G$ be  positive. However, in this case a
negative mass graviton solution arises assuming standard boundary conditions. It was observed in \cite{comgrav1} that if the topological mass $\mu$ is related to the cosmological constant $\Lambda$ by $\mu=\sqrt{-\Lambda}$, and suitable boundary conditions  are imposed, then this negative mass graviton mode can be 
confined to propagate only on the boundary.\footnote{There has also been
an alternative approach in which the bulk graviton is maintained but 
the negative
energy black hole solution is viewed as being possibly irrelevant 
by imposing a suitable superselection rule \cite{cadewawi}.}
 It would be interesting to study the  properties of chiral TMG at the quantum level. This is complicated by the fact that there is an enhancement of the local symmetries at the chiral point  \cite{proof}. One can ask instead whether a generic TMG, upon quantization, flows to the chiral point. To this effect the one-loop beta functions for the dimensionless couplings $\widetilde G= Gk$, $\widetilde\Lambda=\Lambda/k^2$ and $\tilde\mu=\mu/k$, where $k$ is the cut-off parameter, have  been computed in \cite{Percacci:2010yk} for generic values of the couplings.
It was found that the one-loop beta function for $\nu \equiv \mu G=\tilde \mu\widetilde G$
(the coefficient of the Chern-Simons term) vanishes. Then the RG flow occurs in the $\widetilde\Lambda$-$\widetilde G$ plane with  $\nu$ held constant. This two-dimensional flow was shown not to preserve the ratio $\mu^2/\Lambda=\tilde\mu^2/\widetilde\Lambda^2$.

In this paper we shall study the one-loop beta functions in the locally supersymmetric version of TMG, which we shall refer to as TMSG. Our principal motivations for doing so are as follows. Firstly, the determination of whether local supersymmetry helps in making the chiral point condition robust upon the running of the coupling constants.

Another motivation comes from studies of the renormalization group for gravity 
\cite{reuter1}, mostly with the aim of supporting the hypothesis of 
asymptotic safety \cite{weinberg,reviews}.
Most of this work has been done in gravity, possibly coupled to ordinary matter, in four dimensions.
\footnote{In the so-called Einstein-Hilbert truncation the results seems to be relatively independent of dimension, but when one looks in detail at the the physical mechanism underlying the existence of the nontrivial fixed point there are interesting differences
above and below three dimensions \cite{nink}.}
In this work we shall extend this approach to supergravity, also taking into account the gravitational Chern-Simons term, with the attendant subtleties due to the odd number of derivatives in the field equation.

Finally, we wish to develop methods to deal with the renormalization group analysis in three-dimensional supergravities, which apparently have not been addressed so far in the literature. There are a number of subtleties having to do with the fact that the Chern-Simons term has an odd number of derivatives, with the dependence on gauge conditions and on cut-off schemes.
Here we have developed methods which can be applied in a wider class of theories.
In particular, we use the proper time flow equation \cite{flop},
combined with a simple choice of cutoff, to express the beta functions directly
in terms of the heat kernels of appropriate wave operators.

Our main finding with regard to the fate of the chirality condition is that local supersymmetry does not qualitatively change the conclusion reached in the purely bosonic TMG. The general structure of the flow is not altered significantly by the presence of the fermionic fields: for fixed $\nu$ the flow in the $\tilde\Lambda$-$\widetilde G$ plane has a
Gaussian fixed point (at vanishing couplings) with one UV-attractive and one repulsive direction, and a non-Gaussian fixed point with positive $\widetilde G$ which is UV-attractive in both directions.

The paper is organized as follows. In section 2 we describe the theory.
In section 3 we describe the method used to compute the beta functions.
In section 4 we give the expansion of the action to second order in fluctuations.
In section 5 we give the calculation of the beta functions for pure supergravity,
{\it i.e.} in the absence of Chern-Simons term.
The calculation of the beta functions for TMSG is given in section 6
and the corresponding flows are described in section 7.
Section 8 contains final comments and conclusions.
Several helpful formulae and computations have been relegated to appendices A-F.


\section{Topologically Massive Supergravity}


The action for topologically massive off-shell $N=1$ supergravity is given 
by\footnote{This is a straightforward generalisation \cite{rocvan} 
of the on-shell
model of Deser and Kay \cite{deskay}, and its extension by Deser \cite{deser}
to include the cosmological constant.  The pure off-shell supergravity with
cosmological constant was constructed in superspace in \cite{gagrrosi}.}
\begin{eqnarray}
\label{tmsg}
e^{-1} {\cal L} &=& Z\bigl[R  -2 S^2 - 4mS - 2 \varepsilon^{\mu\nu\rho} {\bar\psi}_\mu
D_\nu (\omega) \psi_\rho - m {\bar\psi}_\mu \gamma^{\mu\nu} \psi_\nu
\nn\w2
&& -\ft14 \mu^{-1}\,\varepsilon^{\mu\nu\rho}\left( R_{\mu\nu}{}^{ab} \omega_{\rho ab}
+ \ft23 \omega_\mu^{ab} \omega_{\nu b}{}^c\omega_{\rho ca} \right)
-\mu^{-1} {\bar R}^\mu \gamma_\nu\gamma_\mu R^\nu\bigr]\ ,
\end{eqnarray}
where $Z=\frac{1}{16\pi G}$, $m = \sqrt{-\Lambda}$ and the curvatures are given by
\bea
R_{\mu\nu}{}^{ab} &=& \partial_\mu
\omega_\nu^{ab}+ \omega_\mu^{ac} \omega_{\nu c}{}^b- (\mu
\leftrightarrow \nu)\ ,\w2
R^\mu &=& \varepsilon^{\mu\nu\rho}
 D_\nu(\omega) \psi_\rho\ .
\eea
The real scalar $S$ is the auxiliary field and the covariant derivative of the gravitino in \eq{1} is defined
as $D_{[\mu}(\omega)\psi_{\nu]} = \partial_{[\mu} \psi_{\nu]} +\ft14 \omega_{[\mu}{}^{ab} \gamma_{|ab|} \psi_{\nu]}$.
The spin connection is not an independent field, but rather it is given by
\be
\omega_{\mu ab} = \omega_{\mu ab}(e) + \ft12\left( {\bar\psi}_\mu\gamma_a\psi_b -{\bar\psi}_\mu\gamma_b\psi_a +
{\bar\psi}_a\gamma_\mu\psi_b\right)\ ,
\label{ct}
\ee
where  $\omega_{\mu ab}(e)$ is the spin connection that solves the vanishing torsion equation
$de^a+\omega^a{}_b \wedge e^b=0$. The action is invariant under the local supersymmetry transformations \cite{uematsu}
\bea
\delta e_\mu^a &=& {\bar\epsilon}\gamma^a\psi_\mu\ ,
\nn\w2
\delta \psi_\mu &=& D_\mu(\omega)\epsilon + \ft12 S \gamma_\mu \epsilon\ ,
\nn\w2
\delta S &=& \frac12 {\bar\epsilon}\gamma^\mu R_\mu -\frac12 {\bar\epsilon}\gamma^\mu\psi_\mu S\ .
\label{susy}
\eea
The field equation for $S$ gives $S=-m$. Substituting this back into the action yield the on-shell theory
with the Lagrangian \cite{deskay,deser}
\begin{eqnarray}
e^{-1} {\cal L} &=& Z\bigl[R  +2m^2 - 2 \varepsilon^{\mu\nu\rho} {\bar\psi}_\mu
D_\nu (\omega) \psi_\rho - m {\bar\psi}_\mu \gamma^{\mu\nu} \psi_\nu
\nn\w2
&& -\ft14 \mu^{-1}\,\varepsilon^{\mu\nu\rho}\left( R_{\mu\nu}{}^{ab} \omega_{\rho ab}
+ \ft23 \omega_\mu^{ab} \omega_{\nu b}{}^c\omega_{\rho ca} \right)
-\mu^{-1} {\bar R}^\mu \gamma_\nu\gamma_\mu R^\nu\bigr]\ ,
\label{1}
\end{eqnarray}
and supersymmetry transformations
\bea
\delta e_\mu^a &=& {\bar\epsilon}\gamma^a\psi_\mu\ ,
\nn\w2
\delta \psi_\mu &=& D_\mu(\omega)\epsilon - \ft12 m \gamma_\mu \epsilon\ .
\label{susy2}
\eea
The maximally supersymmetric vacuum solution is given by the AdS$_3$ metric ${\bar g}_{\mu\nu}$ with curvature scalar ${\bar R}=-6m^2$.


\section{The Method For Computing the Beta Functions}


\subsection{Proper Time Representation of the Beta Functions}


In this section we describe the general idea behind the calculational method we shall use.
The one-loop effective action can be written formally as\footnote{For fermions the formula is
$\Gamma=S-\tr\log(\Delta_F)$.}
\begin{equation}
\Gamma=S+\frac{1}{2}\tr\log(\Delta)\ ,
\end{equation}
where $S$ is the classical action and $\Delta=\frac{\delta^2 S}{\delta\phi^2}$, the inverse propagator,
is a differential operator of dimension\footnote{Usually $\omega$ is also equal to the order of the differential
operator, but in this paper we will need to distinguish the two notions.} $\omega$ with eigenvalues $\lambda_n$
and multiplicities $d_n$. We implicitly assume that spacetime is compact without boundary. The trace of the
logarithm can be written, again formally, in the proper time representation
\begin{equation}
 \tr\log(\Delta) = \log\det\Delta = -\int_0^\infty \frac{dt}{t}\,Y(t)\ ,
\end{equation}
where
\begin{equation}
\label{hk}
Y(t)=\sum_n d_n e^{-t\lambda_n}
\end{equation}
is the trace of the heat kernel of $\Delta$. Note that the dimension of $t$ is $-\omega$.
The lower end of the integration corresponds to the UV, the upper end to the IR. One can make sense of this expression by cutting off the integral over small $t$. We also cut-off the integral for large $t$, thereby eliminating any spurious IR divergences. Ignoring the UV problems for a moment, we define the Wilsonian one-loop effective action $\Gamma_k$ as \footnote{Due to the presence of the cut-off function, $\Gamma_k$ is no longer a product or ratio of determinants, as explained in appendix E.}
%
\begin{equation}
\label{hkrep}
\Gamma_k=S-\frac{1}{2}\int_0^\infty \frac{dt}{t}\,Y(t) C_k(t)\ .
\end{equation}
Here $C_k(t)$ is a dimensionless cutoff function which can be written as $C_k(t)=\tilde C(\tilde t)$, where $\tilde t=t k^\omega$ and $\tilde C$ itself does not depend on $k$. The function $\tilde C$ is required to be monotonically decreasing; to go rapidly to zero for $\tilde t\gg 1$; and for $\tilde t\ll 1$ $\tilde C$ should to go sufficiently rapidly to one \cite{flop}. The functional $\Gamma_k$ contains the contribution of all quantum fluctuations with momenta larger than $k$, and therefore it can be regarded as a realization of the Wilsonian prescription for an ``effective action'' at scale $k$. We can define a ``beta function'' of the theory as the logarithmic derivative of $\Gamma_k$:
\begin{equation}
\beta=k\frac{d\Gamma_k}{d k}=-\frac{1}{2}\int_0^\infty \frac{dt}{t}\,Y(t)\,
k\frac{dC_k(t)}{d k}\ .
\label{betafunctional}
\end{equation}
Owing to the fall-off properties of $C_k$, this ``proper time beta function'' is automatically UV convergent, even though the functional $\Gamma_k$ itself is ill-defined in the UV. In fact, the integral receives its main contribution from momenta of order $k$. One can therefore take the view that $\beta$ is the basic object and that $\Gamma_k$ can be obtained by integrating the flow defined by $\beta$.

The beta functions of individual couplings in $\Gamma_k$ can be obtained as the coefficients of the respective operators in the functional $\beta$. The common way of calculating approximate beta functions is to truncate the effective action to contain only the terms of interest. For example, to obtain the beta functions of $\Lambda$, $G$ and $\mu$ one can assume that the effective action has the form \eq{1} and use it to calculate the r.h.s. of \eq{betafunctional}.

The beta functions obtained in this way will generally depend on the choice of the cutoff function $C_k(t)$. We shall refer to this as {\it scheme dependence}. However, the beta functions of the dimensionless couplings are scheme-independent. This can be seen as follows. Let $Y_n$ be the coefficient of $t^n$ in the series expansion of $Y$. In particular $Y_0$, the $t$-independent term, is dimensionless, so its coefficient in the action is a dimensionless coupling. Using the homogeneity and the boundary conditions of $\tilde C$ we have
\begin{equation}
-\frac12\int_0^\infty \frac{dt}{t}\,Y_0\,k\frac{dC_k(t)}{d k}
=-\frac12 \omega\, Y_0\int_0^\infty d\tilde t\, \frac{d\tilde C}{d \tilde t}
=\frac12 \omega\, Y_0\ .
\end{equation}
Thus we see that the beta functions of dimensionless couplings are actually ``universal'' in the sense that they do not depend on the choice of cutoff function. When the flow equation is integrated, these couplings run logarithmically, and in the limit $k\to\infty$ they correspond to logarithmic
divergences in $\Gamma_k$. On the other hand, the beta function of the coupling that
multiplies the term $Y_n$ ($n\not=0$) will be scheme dependent. It will scale as $k^{-n\omega}$ and therefore, for $n<0$, corresponds to a power law divergence. These beta functions coincide with those that one would obtain as the coefficients of divergent terms in $\Gamma$.

\subsection{Theta Function Cutoff}\label{thetacutoffsec}

Let us consider the cutoff
\begin{equation}
\tilde C(\tilde t)=\theta(1- a \tilde t)\ ,
\label{c1}
\end{equation}
where $\theta$ is the Heaviside step function, $a$ is a constant parameter we have introduced, and
we recall that $\tilde t=t k^\omega$. Then
\begin{equation}
k\frac{dC_k(t)}{dk}
=-a\omega \tilde t\,\delta(1-a\tilde t)
=-\omega t\,\delta\left(\frac{k^{-\omega}}{a}-t\right)\ .
\end{equation}
When we insert this in (\ref{betafunctional}) we get simply
\begin{equation}
\label{genbeta}
\beta\equiv k\frac{d\Gamma_k}{dk}=\frac12 \omega\, Y\left(\frac{k^{-\omega}}{a}\right)\ .
\label{c2}
\end{equation}
In this regularization scheme the one loop beta functions of the individual couplings
can be simply obtained from the small-$t$ expansion of the heat kernel $Y(t)$,
for which much information is available in the literature.
An alternative choice of cutoff that also allows an explicit evaluation
of the beta functions is discussed in appendix B.


\subsection{The Evaluation of the Heat Kernel}


In this paper we will have to evaluate the heat kernel for differential operators
$\Delta_1, \Delta_2$ and $\Delta_3$ of order $1, 2$ and $3$ respectively.
Assuming that the coefficients of the highest order terms are dimensionless,
the corresponding kernels are
\be
Y_1(u)={\rm tr}\, e^{-u\Delta_1}\ ,\qquad
Y_2(t)={\rm tr}\, e^{-t\Delta_2}\ ;\qquad
Y_3(s)={\rm tr}\, e^{-s\Delta_3}\ ,
\label{stu}
\ee
where $u$, $t$ and $s$ are real parameters of dimension $L$, $L^2$ and
$L^3$, respectively. In the following, we will encounter situations where the highest order part
of the operator is multiplied by $1/\mu$. By expanding the exponential for
small or large $\mu$, we will reduce the calculation to the evaluation of traces
of the form given above with insertions of operators coming from the $\mu$ expansions.
Such traces will be dealt with in the same way as we shall now describe.

The evaluation of the sums $Y(t)=\sum_n d_n e^{-t\lambda_n}$ can be conveniently carried out by using the Euler-Maclaurin formula,
\be
\sum_{n=n_0}^\infty F(n,t) = \int_{n_0}^\infty F(x,t)\, dx -\sum_{k\ge 0}
\fft{B_{k+1}}{(k+1)!}\, F^{(k)}(n_0,t)\ ,
\label{euma}
\ee
where $F(x,t)=d_x e^{-t\lambda_x}$ and $B_k$ is the $k$'th Bernoulli number. Note that since we need only the terms in the small-$t$ expansion of $Y(t)$ up to and including the $t^0$ term, only the first few terms in the summations involving the Bernoulli numbers will be required. Since the terms in the summation can only contribute non-negative powers of $t$, in our calculation they only appear in the $t$-independent terms.
The integral has the asymptotic expansion
\be
\int_{n_0}^\infty F(x,t)dx=I_{-3/2}\, t^{-3/2}+I_{-1}\, t^{-1}
+I_{-1/2}\, t^{-1/2}+I_0+O(t^{1/2})\,.\label{FI}
\ee
The resulting spectral sums can be expanded in powers of Ricci scalar $R$.
The leading terms are $R$-independent and they are given by
\be
\label{coefs}
Y^{(1)}(u) = \frac{VN_1}{\pi^2 u^3}\ ,\qquad
Y^{(2)}(t) = \frac{VN_2}{(4\pi t)^{3/2}}\ ,\quad
Y^{(3)}(s) = \frac{VN_3}{6\pi^2 s}\ ,
\label{uts}
\ee
where $N_i$ are the numbers of independent components of the field on which the operators act.
The beta functions will consist of appropriately weighted sums of the heat kernels.
There is freedom in introducing a suitable proportionality factor in the relations between $u,t,s$ and $k$.
This can be viewed as another instance of {\it scheme-dependence}. We will choose
\be
t=u^2 \pi^{1/3}/4\ ,\qquad s=u^3/6\ ,
\label{ts}
\ee
in such a way that the denominators in \eq{uts} become equal so that $Y^{(i)}(t)= N_iV/(4\pi t)^{3/2}$.
We show in appendix D that these choices are natural, since they imply that the
leading terms are the same when the beta functions are computed directly from the
heat kernel of the Dirac operator or from the heat kernel of its square.

\subsection{Beta Function Definitions for Topologically Massive \hfill\break Supergravity}

The beta function of the theory, being expressible in terms of heat kernels, will have the same general structure as the heat kernels themselves. When evaluated on a Euclidean AdS (i.e. $S^3$; see appendix C) background, it will have the form
\bea
k\, \fft{d\Gamma_k}{dk} &=& \frac{V k^3}{16\pi}
\Big[A(\tilde\Lambda,\tilde\mu)
+ B(\widetilde\Lambda,\tilde\mu)\tilde R
+ C(\widetilde\Lambda,\tilde\mu)\tilde R^{3/2}+O(\tilde R^2)\Big]\ ,
\label{mr}
\eea
where we have inserted powers of $k$ such that the coefficients $A,B$ and $C$, and the tilded quantities
\be
\widetilde\Lambda = \frac{\Lambda}{k^2}\ ,\qquad \widetilde\mu = \frac{\mu}{k}\ ,\qquad \widetilde R= \frac{R}{k^2}
\ee
are dimensionless. The prefactor $1/(16\pi)$ is conventional and is useful to simplify the form of the beta functions. The volume of $S^3$ with radius $\ell$ is $V(S^3)=2\pi^2 \ell^3$ with $\ell= \sqrt{\frac6{R}}$.

Evaluating the Euclidean version of the renormalized TMSG action \eq{1} on the
$S^3$ background, it can be written in the form
\be
\Gamma_k = V \left(\frac{ 2\Lambda}{16\pi G}- \frac{1}{16\pi G} R
+\frac{1}{12 \sqrt 6 \pi G \mu} R^{3/2}+ O(R^2)\right)\ ,
\label{gs}
\ee
where we have used that the integral of the  CS term on $S^3$ is given by $\int \tr (\omega d\omega +\frac23 \omega^3)=32 \pi^2$. The couplings $\Lambda$, $G$, $\mu$ are now renormalized couplings evaluated at scale $k$.
In addition, rescaling the coupling constant $G$ as
\be
\label{rescalingcouplings}
G = \widetilde G k^{-1}\ ,
\ee
so as to make $\widetilde G$ dimensionless, and comparing the $t$-derivative of \eq{gs} with \eq{mr}, we obtain:
\bea
k\, \fft{d\tilde\Lambda}{dk}
-k\, \fft{d\tilde G}{dk}\, \fft{\widetilde G}{\widetilde\Lambda} &=&
- 3\tilde\Lambda+\frac{1}{2}A\tilde G\ ,
\\
k\, \fft{d\tilde G}{dk} &=& \widetilde G
+B\tilde G^2\ ,
\\
\frac{1}{\tilde\mu\tilde G}
\left(k\, \fft{d\tilde G}{dk}\, \widetilde G^{-1}
  +k\, \fft{d\tilde \mu}{dk}\, \tilde \mu^{-1}\right) &=& -\frac{3\sqrt3}{2\sqrt2}C\ ,
\label{beta}
\eea
From the first two equations one obtains the one-loop beta functions of $\widetilde G$ and $\widetilde\Lambda$:
\bea
k\, \fft{d\tilde G}{dk} &=&\widetilde G+B\,\widetilde G^2\ ,
\nonumber
\\
k\, \fft{d\tilde\Lambda}{dk}  &=& -2\tilde\Lambda
+\frac{1}{2}A\tilde G+B\tilde G\tilde\Lambda \ .
\label{oneloopbeta}
\eea
These equations have exactly the same form as in pure gravity with
cosmological constant, except that the coefficients $A$ and $B$
will depend on $\tilde\mu$.
From equation \eq{beta} one can determine the running of $\mu$.


\section{The Quadratic Action and Spectra }


The approach we shall take is to Euclideanize the theory,
and consider the special case of a 3-sphere background \cite{Deger:1998nm}.
(The rules for Euclideanization are summarized in appendix C.)
In this background, we can write down the eigenvalues of all
the relevant operators describing the quadratic fluctuations of the action,
and then perform the sums in (\ref{hk}).  By making use of the
Euler-Maclaurin summation formula, we are able to obtain asymptotic
expansions for the $Y(t)$ functions for the various operators.


\subsection{The Bosonic Sector}


The first step is to calculate the operator ${\mathcal{O}}$ that describes
the quadratic fluctuations of the action:
\be
S^{(2)}_h =\frac{Z}{4} \int d^3x\, \sqrt{-g}\,
h_{\mu\nu}\, {\cal O}^{\mu\nu,\rho\sigma}\, h_{\rho\sigma}\,.
\ee
In the metric formalism, it can be read off from eq. (3.7) of ref. \cite{Percacci:2010yk}.
Since we are considering a theory that contains spinor fields we must work in dreibein formalism, and this gives rise to a new contribution to ${\cal O}$, which can be understood as follows.
The first variation of the action in the metric formalism is of the form
$\delta g_{\mu\nu}E^{\mu\nu}$, where $E^{\mu\nu}=G^{\mu\nu}+\Lambda g^{\mu\nu}+\frac{1}{\mu}C^{\mu\nu}$,
$G^{\mu\nu}$ being the Einstein tensor and $C^{\mu\nu}$ the Cotton tensor. The second variation is then obtained by varying $E^{\mu\nu}$. In the dreibein formalism the first variation is
$\eta_{ab}\delta e^a_{(\mu}e^b_{\nu)}E^{\mu\nu}$. The second variation contains, in addition to the variation of $E^{\mu\nu}$ also a term $\eta_{ab}\delta e^a_{(\mu}\delta e^b_{\nu)}E^{\mu\nu}$.
This term vanishes on shell, but since we are calculating the beta functions off shell, it has to be retained \cite{harst,dona}.
Since the Cotton tensor is proportional to covariant derivatives of the Ricci tensor and Ricci scalar, it vanishes for the metric of the sphere. Therefore the additional terms in the second variation are just
\be
\frac{Z}{24} \int d^3x\sqrt{-g}\,(6\Lambda-R)h_{\mu\nu}h^{\mu\nu}\ ,
\label{new}
\ee
where $h_{\mu\nu}=2e_{a\mu}\delta e^a_\nu$.

Since AdS$_3$ (and $S^3$) have no moduli, the resulting operator ${\cal O}$ has zero modes only corresponding to infinitesimal coordinate transformations and local Lorentz transformations. To make it invertible, one adds the coordinate gauge fixing term
\be
S_{GF}^B =  -\frac{Z}{2\a}\int d^3 x \sqrt{-\bar g} G_\mu {\bar g}^{\mu\nu} G_\nu\ ,
\label{gfa}
\ee
where\footnote{We use the convention that $D_\mu$ is the covariant derivative using the spin connection whereas $\nabla_\mu$ means covariant derivative using the Christoffel symbol.}
\be\label{G}
G_\nu =\nabla_\mu h^\mu{}_\nu -\frac{\beta + 1}{4} \partial_\nu h
\ee
Then one has to add the ghost action
\be
S_{gh}^B = \int d^3 x \sqrt {-g}\,\,{\bar C}^\mu\!
\left( -\delta_\mu^\nu \Box
-\frac{1-\beta}{2} \nabla_\mu\nabla^\nu
-R_\mu{}^\nu  \right) C_\nu\ ,
\label{gh}
\ee
where $C_\mu$ is an anticommuting complex vector. A standard gauge condition to fix the local Lorentz symmetry is to set the antisymmetric part of the dreibein equal to zero \cite{VanNieuwenhuizen:1981ae}. This leads to a ghost Lagrangian of the form $\bar C_{ab}( C^{ab}+D^a C^b)$ where $C_a=C_\mu \bar e^\mu_a$ is the ghost associated with the general coordinate transformations.
Redefining $C^{ab}+D^a C^b=C^{\prime ab}$ we see that the ghost $C^{\prime ab}$ does not propagate and hence it will be neglected.

In order to extract the eigenvalues of the operator ${\mathcal{O}}$
it is convenient to decompose the graviton field $h_{\mu\nu}$
into its irreducible parts: the spin-2 transverse traceless part $h_{\mu\nu}^{TT}$,
the spin-1 transverse vector $\xi^{T\mu}$, the spin-0 components $\sigma$ and $h$:
\be
h_{\mu\nu}=h^{TT}_{\mu\nu}+\nabla_\mu\xi^T_\nu+\nabla_\nu\xi^T_\mu
+\nabla_\mu\nabla_\nu\sigma-{1\over3}g_{\mu\nu}\Box\sigma+{1\over3}
g_{\mu\nu}h\ .
\label{decomp}
\ee
Similarly, the ghost is decomposed into a spin-1 transverse vector $V$
and a scalar $S$:
\be
C_\mu=V_\mu+\nabla_\mu S\,.
\ee
It is also convenient to define
\be
\sqrt{-\Box-\frac{R}3 }\ \xi^T_\mu  = \xi'^T_\mu\ ,\qquad
\sqrt{(-\Box)\left(-\Box-\frac{R}2\right)}\ \sigma = {\sigma'}\ , \qquad
\sqrt{-\Box} S = {S'}
\ee
The Jacobian of this field redefinition cancels the one of \eq{decomp}.


\subsubsection{The Diagonal Gauge}


In the following we restrict ourselves to the ``diagonal'' gauge
\be
\beta=\fft{(2\alpha+1)}{3}
\ee
which will ensure that there is no mixing between $\sigma$ and $h$. At this point we pass to the Euclidean theory (see appendix C).
The quadratic part of the Euclideanized bosonic action reads
\be
\label{s2}
S^{(2)}+S_{GF}^B =\frac{Z}{4} \int d^3 x \sqrt{g}\left[
h^{TT\mu\nu } \Delta_{(h^{TT})\mu\nu}{}^{\rho\sigma} h_{\rho\sigma}^{TT}
+ c_\xi {\xi'}^{T\mu} \Delta_{(\xi^T)\mu}{}^\nu \xi'^T_\nu
+ c_\sigma {\sigma'}\Delta_{(\sigma)} {\sigma'}
+ c_h h \Delta_{(h)} h \right]\ ,
\ee
and the ghost action reads
\be
\label{gh2}
S^{B}_{\rm ghost} = \int d^3 x \sqrt {g} \left[
{\bar V}^\mu \Delta_{(V)\mu}{}^\nu V_\nu
+c_S\bar{ S'} \Delta_{(S)} {S'} \right]\ ,
\ee
where we have defined the operators \cite{Percacci:2010yk}\footnote{In comparing with \cite{Percacci:2010yk}, one needs to take into account the new contribution \eq{new} which arises due to the use of the dreibein formalism.}
\bea
\label{bops}
\Delta_{(h^{TT})\mu\nu}{}^{\rho\sigma} &=&
\left(-\Box+\frac{R}{2}-\Lambda\right) \delta_{(\mu}^{(\rho} \delta_{\nu)}^{\sigma)}
+\frac{1}{\mu} \varepsilon_{(\mu}{}^{\lambda(\rho}\delta_{\nu)}^{\sigma)}
\nabla_\lambda\left(\Box -\frac{R}{3}\right) \ ,
\nn\w2
\Delta_{(\xi^T)\mu}{}^\nu &=&
\left(-\Box- \frac{3\alpha-2}{6} R -3\alpha\Lambda \right)\delta^\nu_\mu\ ,
\nn\w2
\Delta_{(\sigma)} &=&
-\Box -\frac{R}{2} -\frac{3\alpha\Lambda}{2(4-\alpha)}\ ,
\nn\w2
\Delta_{(h)} &=&
-\Box -\frac{12\Lambda}{4-\alpha}\ ,
\nn\w2
\Delta_{(V)\mu}{}^\nu &=& \left(-\Box-\frac{R}{3}\right)\delta_\mu^\nu\ ,
\nn\w2
\Delta_{(S)} &=& -\Box-\frac2{4-\alpha} R\ ,
\eea
and coefficients
\be
c_\xi = \frac{2}{\alpha}\ ,\qquad
c_\sigma = \frac{2(4-\alpha)}{9\alpha}\ ,\qquad
c_h = -\frac{4-\alpha}{18}\ ,\qquad
c_S=\frac{4-\alpha}{3} \ .
\label{cs1}
\ee
Using the results of \cite{Percacci:2010yk}, the eigenvalues of these operators are found to be
\bea
\lambda_n^{h^{TT\pm}} &=& \kur^2(n^2+2n+1)-\Lambda\pm \frac{\kur^3}{\mu} n(n+1)(n+2)\ ,\quad n\ge 2\ ,
\nn\\
\lambda^{\xi^T}_n &=& \kur^2\left(n^2+2n-3+3\alpha\right) -3\alpha\Lambda\ , \quad n\ge 2\ ,
\nn\\
\lambda^\sigma_n  &=& \kur^2\left(n^2+2n-3\right)-\frac{3\alpha\Lambda}{4-\alpha} \ , \quad n\ge 2\ ,
\nn\\
\lambda^h_n &=& \kur^2\left(n^2+2n\right)-\frac{12\Lambda}{4-\alpha}\ ,\quad n\ge 0\ ,
\nn\\
\lambda^V_n &=& \kur^2\left( n^2 + 2n -3\right)\ ,\quad n\ge 1\ ,
\nn\\
\lambda^S_n &=& \kur^2\left(n^2 + 2n - \frac{12}{4-\alpha}\right)
\ ,\quad n\ge 1\ ,
\label{beigen}
\eea
where we have defined
\be
\kur \equiv \sqrt{\fft{R}{6}}
\ee
and the multiplicities are
\bea
\label{multiplicities}
d^{T+}_n= d^{T-}_n &=& n^2+2n-3\ ,\nn\\
d^\xi_n=d_n^V&=& 2(n^2+2n)\ ,\nn\\
d^\sigma_n=d_n^h=d_n^S&=& n^2+2n+1\ .
\eea

Requiring positivity of the Euclideanized version of the gauge fixing action \eq{gfa},
and staying on one side of the singular point $\alpha=4$, we are led to impose the condition
\be
0\le\alpha<4\ .
\label{alpharange}
\ee
Then, $c_h<0$ and the operator $\mathcal{O}$ acting on the trace $h$ is negative. This corresponds to the well-known conformal factor problem \cite{Gibbons:1978ac}. The $\alpha=0$ case is special and it will be discussed next.

\subsubsection{The Physical Gauge}

It is sometimes convenient to use a slightly different approach to quantisation, in which one works in a physical gauge rather than integrating also over the gauge degrees of freedom.  In the present context, this amounts to
setting to zero, as a physical gauge choice, the longitudinal part of the metric fluctuations, which correspond to general coordinate transformations.  In our notation, this means that $\xi_\mu^T$ and $\sigma$ should be set to zero. This can be accomplished as follows. Setting $\alpha=0$ implies that the gauge condition $\nabla_\mu h^\mu{}_\nu -\frac{\beta + 1}{4} \partial_\nu h=0$ is to be imposed strongly in the sense that it can be used in the action. Substituting for $h_{\mu\nu}$
\be
h_{\mu\nu}=h^{TT}_{\mu\nu}+\nabla_\mu\xi_\nu+\nabla_\nu\xi_\mu +{1\over3} g_{\mu\nu}h'\ .
\label{decomp5}
\ee
where $\xi^\mu$ is no longer divergence-free and $h'$ is no longer the trace of $h_{\mu\nu}$, and choosing $\beta=1/3$, the gauge condition becomes
\be
\nabla^\mu(\nabla_\mu\xi_\nu+\nabla_\nu\xi_\mu) -\frac23 \nabla_\nu \nabla_\mu\xi^\mu = 0\ .\label{confeq}
\ee
Multiplying this equation by $-\xi^\nu$ and integrating over the Euclidean-signature compact manifold without boundary gives
\be
\fft12 \int\sqrt{g}\,d^3x\, (\nabla_\mu\xi_\nu + \nabla_\nu\xi_\mu -\ft23 g_{\mu\nu} \nabla_\rho\xi^\rho)^2=0\,,
\ee
which shows that the kernel of the operator in (\ref{confeq}) is the conformal Killing vectors, which satisfy
\be
 \nabla_\mu\xi_\nu + \nabla_\nu\xi_\mu -\ft23 g_{\mu\nu} \nabla_\rho\xi^\rho=0\,.
\ee
There are in total ten conformal Killing vectors, of which six are Killing vectors, on $S^3$. We can therefore set $\xi^\mu=0$ in the action, and take account of the ten zero modes later, in the computation of the heat kernel. This means setting $\xi^{T\mu}=0, \sigma=0$ and $h'=h$. Since in this gauge one deals only with
the physical degrees of freedom $h_{\mu\nu}^{TT}$ and $h$, we shall call this the ``physical gauge''.
Thus, in the physical gauge the action \eq{s2} becomes
\be
\label{s3}
S^{(2)}+S_{GF}^B = \frac{Z}{4} \int d^3 x \sqrt{g}\Bigl\{
h^{TT\mu\nu } \Delta_{(h^{TT})\mu\nu}{}^{\rho\sigma} h_{\rho\sigma}^{TT}
+\frac29 h (\Box+3\Lambda)h\, \Bigr\}\ .
\ee
Regarding the ghost action, however, setting $\alpha=0$ in \eq{gh2} does not produce the correct answer. Instead, one needs to consider the Jacobian associated with the changing of the path integral measure, namely
\be
{\cal  D}h_{\mu\nu} = Z_{\rm gh} {\cal D}h^{TT}_{\mu\nu} {\cal D}\xi_\mu {\cal D}h'\ ,
\ee
where \cite{Fradkin:1981iu,Bern,Gaberdiel:2010xv}
\be
Z_{\rm gh}= \sqrt{ \mbox{det}_{1}\left(\Box +\frac{R}{3} \right) \mbox{det}_{0}\left(\Box+\frac{R}{2}\right)}\ .
\ee
The Jacobian $Z_{\rm gh}$ can be represented in the path integral by using
\bea
\sqrt{ \mbox{det}_{1}\left(\Box +\frac{R}{3} \right)} &=& \mbox{det}_{1}\left(\Box +\frac{R}{3}\right)
\left(\mbox{det}_{1}\left(\Box +\frac{R}{3}\right)\right)^{-1/2}
\nn\w2
&=&  \int {\cal D} u^\mu {\cal D}v^\mu
\exp \left\{\int d^3 x \left[u^{\mu\star} \left(\Box +\frac{R}{3}\right) u_\mu + v^\mu \left(\Box +\frac{R}{3}\right)v_\mu\right]\right\}\ ,\qquad\qquad
\eea
and similarly
\bea
\sqrt{ \mbox{det}_{0}\left(\Box +\frac{R}{2} \right)} &=& \mbox{det}_{1}\left(\Box +\frac{R}{2}\right)
\left(\mbox{det}_{0}\left(\Box +\frac{R}{2}\right)\right)^{-1/2}
\nn\w2
&=&  \int {\cal D} u {\cal D}v
\exp \left\{\int d^3 x \left[u^* \left(\Box +\frac{R}{3}\right) u + v \left(\Box +\frac{R}{3}\right)v\right]\right\}\ ,\qquad\qquad
\eea
where $(u^{\mu},u)$ are {\it anticommuting} complex vector and scalar fields and $(v^{\mu},v)$ are {\it commuting} real vector and scalar fields. These are Nielsen-Kallosh type ghost fields \cite{nk}.


\subsection{The Fermionic Sector}


We now repeat the steps of the preceding section for the fermions. The first variation of the
fermionic part of the action is given by
\be
S_F^{(1)} = -4Z\int d^3 x \sqrt{-g}\delta{\bar\psi}_\mu \left( R^\mu+\frac12 \gamma^{\mu\nu}\psi_\nu
+\frac{1}{2\mu} C^\mu\right)\ ,
\label{S1}
\ee
where the ``Cottino'' vector-spinor is given by
\be
C^\mu  =  \gamma^\rho \gamma^{\mu\nu} \nabla_\nu R_\rho -\varepsilon^{\mu\nu\rho}
\left(R_{\rho\sigma} -\ft14 g_{\rho\sigma} R \right) \gamma^\sigma \psi_\nu \ ,
\ee
Next we perform the second variation, denoting by $\psi_\mu$ the fluctuation of the gravitino field
without using the background field equations but rather the supersymmetric background
given by the AdS$_3$ metric whose inverse radius $\ell^{-1}$ is not identified with $m$, so as to remain off-shell. Furthermore, decomposing the gravitino field as
\begin{eqnarray}
&&\psi_\mu = \phi_\mu +\left(D_\mu-\frac13\gamma_\mu \slashed D\right)\chi
+\frac13 \gamma_\mu \psi\ ,
\label{decompf}
\w2
&& D^\mu \phi_\mu =0 \ ,\quad \gamma^\mu\phi_\mu=0\ ,
\end{eqnarray}
we find that
\begin{eqnarray}
S^{(2)}_F &=& \int d^3 x \sqrt{-g} \bigg\{ 2\bar\phi_\mu \left[-\slashed D
+\frac12 m+\frac1{\mu} \left(-\Box +\frac38 R\right)\right]\phi_\mu
\nn\w2
&&\!\!\!\!\!\!\!\!\!\!\!\!\!\!\!\!\!\!\!\!\!\!\!
+\frac49 \bar\chi \left(\Box +\frac18 R\right)
\left(\slashed D -\frac32 m\right)\chi
+\frac49 \bar\psi \left(\slashed D +\frac32 m\right)\psi
-\frac89 \bar\psi \left(\Box +\frac18 R\right)\chi \bigg\}\ .
\end{eqnarray}
%

\subsubsection{The Diagonal Gauge}

It is convenient to choose a gauge condition that eliminates the mixing between $\psi$ and $\chi$.
This is achieved by the gauge fixing term  \cite{VanNieuwenhuizen:1981ae}
\begin{equation}
S_{GF}^F = \frac{4}{9\alpha'} \int d^3 x \sqrt{-g}\,
{\bar F}\nk F \ ,
\end{equation}
where $\alpha'$ is a dimensionless gauge fixing parameter,
\be
\nk=\slashed D -\frac{3}{2}\rho\,,
\label{nk}
\ee
and
\be
\label{F}
F =\alpha' \psi+\left(\slashed D +\frac{3}{2}\rho\right) \chi\,.
\ee
The cancellation of the cross term can be seen by noting that acting on a  spin-$\fft12$ field we have
$\left(\slashed D +\frac{3}{2}\rho\right)\left(\slashed D -\frac{3}{2}\rho\right)=\left(\Box +\frac18 R\right)$.
Performing the decomposition \eq{decompf} of the transformation \eq{susy} and taking the $\gamma$-trace and
the divergence, one finds $\delta\psi=\left(\slashed D +\frac32\rho\right)\epsilon$ and
$\left(\Box +\frac18 R\right)(\delta\chi-\epsilon)=0$. Therefore, the fermionic ghost action is given by
\be
S_{gh}^F = \int d^3 x \sqrt{-g}\,
\bar \eta\left[ \alpha'\left(\slashed D -\frac{3}{2}m\right)
+ \left(\slashed D +\frac{3}{2}\rho\right)\right]\eta\ .
\ee
Given that the gauge fixing involves the operator $\nk$ a factor ${\rm det}(\nk)^{-1/2}$ has to be included in the path integral measure to ensure on-shell gauge independence. This can be represented as a Gaussian integration over Nielsen-Kallosh ghost fields \cite{nk}, comprising {\it commuting} Dirac spinor $\omega$ and an {\it anticommuting} Majorana spinor $\gamma$, with action
\be
S_{NK} = \int d^3 x \sqrt{-g}
\left[\bar\omega\nk\omega+\bar\gamma\nk\gamma\right]\ ,
\ee

At this point it is convenient to perform the redefinition
\begin{equation}
\chi' = \sqrt{\Box +\frac18 R}\ \ \chi \ .
\end{equation}
whose Jacobian cancels that of the transformation \eq{decompf}. The total quadratic fermionic action
including the gauge fixing and ghost terms become
\be
S^{(2)}_F+ S_{GF}^F +S_{gh}^F= \int d^3 x \sqrt{-g}
\left[c_\phi\bar\phi_\mu\Delta_{(\phi)}\phi_\mu
+c_\chi\bar\chi'D_{(\chi)}\chi'
+c_\psi\bar\psi D_{(\psi)}\psi
+c_\eta\bar\eta D_{(\eta)}\eta
 \right]\ ,
 \label{ferma}
\ee
where
\bea
\Delta_{(\phi)}&=&\slashed D-\frac12 m -\frac{1}{\mu} \left(\Box -\frac38 R\right) ,
\nn\\
D_{(\chi)}&=&\slashed D+\frac{3(\rho-\alpha' m)}{2(1+\alpha')} \ ,
\nn\\
D_{(\psi)}&=&\slashed D +\frac{3( m- \alpha'\rho)}{2(1+\alpha')}\ ,
\nn\\
D_{(\eta)}&=&\slashed D+\frac{3(\rho- \alpha' m)}{2(1+\alpha')}\ ,
\label{fops}
\eea
and
\be
c_\phi=-2\ ;\qquad
c_\chi=c_\psi= \frac{4(1+\alpha')}{9\alpha'}\ ;\qquad
c_\eta=1+\alpha' \ .
\label{cs2}
\ee

Note also that the value $\alpha'=-1$ is singular. Thus we shall restrict $\alpha'$ to obey
\be
\alpha' >-1\ ,
\label{betarange}
\ee
which can be seen to be an acceptable range.

Next, we continue from AdS$_3$ to $S^3$ as explained in appendix C, and perform harmonic expansions on $S^3$. The eigenvalues of the Dirac and Laplace operators on the appropriate spinor harmonics on $S^3$ are
\begin{eqnarray}
i\slashed D Y_a^{(\ell,\pm 3/2)} &=& \pm \rho(\ell+1)  Y^{(\ell,\pm 3/2)}\ , \quad\quad\quad \ \ell=\ft32,\ft52,...
\nn\w2
-\Box Y_a^{(\ell,\pm 3/2)} &=& \rho^2\left[\ell(\ell+2) -\ft32 \right] Y_a^{(\ell,\pm 3/2)}\ , \quad \ \ell=\ft32,\ft52,...
\nn\w2
i\slashed D Y^{(\ell,\pm 1/2)} &=& \pm \rho (\ell+1)  Y^{(\ell,\pm 1/2)}\ , \quad\quad\quad \ \ell=\ft12, \ft32...
\nn\w2
-\Box  Y^{(\ell,\pm 1/2)} &=& \rho^2\left[\ell(\ell+2) -\ft12 \right] Y^{(\ell,\pm 1/2)}\ , \quad  \ \ell=\ft12, \ft32...
\label{eigenvalues2}
\end{eqnarray}
with multiplicities $\ell(\ell+2)- \frac54$ for spin 3/2 and $\ell(\ell+2) +\frac34$ for spin 1/2.
Using the formula (\ref{eigenvalues2}) we find, after defining $\ell=n- \frac12$, that the eigenvalues of
the operators listed in \eq{fops} and \eq{nk}  are
\begin{eqnarray}
\lambda_n^{\phi\pm} &=&  \pm \rho(n+\ft52) -\ft12 m +\frac{\rho^2}{\mu}(n+2)(n+3)\ ,
\qquad n=0,1,...
\nn\w2
\lambda_n^{\chi'\pm} &=& \pm \rho(n+\ft32)+\frac{3(\rho-\alpha' m)}{2(1+\alpha')}\ ,
\qquad n=1,2,...
\nn\w2
\lambda_n^{\psi\pm} &=& \pm \rho(n+\ft32)+\frac{3(m-\alpha'\rho)}{2(1+\alpha')}\ ,
\qquad n=0,1,...
\nn\w2
\lambda_n^{\eta\pm} &=& \pm \rho(n+\ft32)+\frac{3(\rho-\alpha' m)}{2(1+\alpha')} ,
\qquad \  n=0,1,...
\nn\w2
\lambda_n^{NK\pm} &=& \pm \rho (n+\ft32)-\frac32 \rho \ ,\qquad n=0,1,...
\label{feigen}
\end{eqnarray}
with multiplicities
\begin{eqnarray}
\label{fmult}
d_{(n,3/2)} &=& (n+1)(n+4)\ ,
\nn\w2
d_{(n,1/2)} &=& (n+1)(n+2)\ .
\end{eqnarray}
Note that for $\lambda_n^{\chi'\pm}$ we leave out the eigenvalues $n=0$
which correspond to Killing spinors and do not contribute to $\psi_\mu$.

\subsubsection{The Physical Gauge}

Letting $\alpha' \to 0$ implies that the gauge condition \eq{F} is to be strongly imposed in the sense that it is to be used in the action. This implies that $ (\slashed{D}+\frac32\rho)\chi=0$, and consequently, $\chi=0$ except for those that are Killing spinors.  Next, it is convenient to decompose $\psi_\mu$ as
\be
\psi_\mu = \phi_\mu + \left(D_\mu -\frac12 m \gamma_\mu\right) \zeta +\frac13 \gamma_\mu \psi'\ ,
\label{decomp3}
\ee
since $\zeta$ will not appear in the action due to the fact that the $\zeta$ dependent term in \eq{decomp3} is a supersymmetry transformation. Comparing the trace of $\psi_\mu$ using \eq{decompf} and \eq{decomp3} we find that
\bea
\left(\Box+\frac{R}{8} \right) (\chi-\zeta)&=& 0\ ,
\label{r1}\w2
\psi -\psi'&=& \left(\slashed{D} -\frac32 m\right)\zeta \ .
\eea
From \eq{r1} it follows that $\chi=\zeta$ up to linear combination of conformal Killing spinors. This can be seen by noting that, acting on a spinor, $\Box+\frac{R}{8}= (\slashed{D}-\frac32\rho)(\slashed{D}+\frac32\rho)$. Thus the physical gauge $\chi=0$ implies that $\zeta=0$ modulo the four conformal Killing spinors of $S^3$, and $\psi=\psi'$. Consequently, in the physical gauge we get
\bea
S^{(2)}_F+ S_{GF}^F = \int d^3 x \sqrt{-g}
\left[-2\bar\phi_\mu\Delta_{(\phi)}\phi_\mu
+\frac49\bar\psi \left(\slashed{D}+\frac32 m \right)\psi
\right]\ .
\label{ferma2}
\eea
In the ghost sector, the correct result is not simply $S_{\rm gh}^F+S_{NK}$ with $\alpha'$ set to zero. Rather, we need to consider the Jacobian associated with the changing of the path integral measure as
\be
{\cal  D}\psi_\mu = Z_{\rm gh} {\cal D} \phi_\mu {\cal D}\zeta {\cal D}\psi' \ ,
\ee
where \cite{Zhang:2012kya}
\be
Z_{\rm gh} = \left[\mbox{det}_{\ft12} \left(\Box +\frac{R}{8}\right)\right]^{-1}\ .
\ee
This admits a path integral representation by using
\be
\left[\mbox{det}_{\ft12} \left(\Box +\frac{R}{8}\right)\right]^{-1}
= \int {\cal D}\kappa \exp \left\{ \int d^3 x \left[ {\bar\kappa}\left( \Box +\frac{R}{8}\kappa\right)\right] \right\}\ ,
\ee
where $\kappa$ is {\it commuting} Dirac spinor field.


\section{The Beta Functions of Pure Supergravity}


The Chern-Simons term in topologically massive supergravity gives rise to a
third-order operator and thus leads to certain complications when
calculating the heat-kernel expansions. In this section we shall therefore begin
by turning off the Chern-Simons term and its superpartners, and consider just
three-dimensional supergravity with a cosmological term.

We have seen in section 3.3 that the beta function of a coupling can be expressed directly
in terms of the heat kernel. In our specific case, each spin component of the graviton and gravitino has a separate heat kernel and we have to specify the way in which these individual contributions
are assembled. For the special case of pure Einstein theory on $S^3$, we show in appendix D that the heat kernel of the complete wave operator $\cal O$ acting on $h_{\mu\nu}$ is reproduced by
simply summing the heat kernels of the individual spin components,
each normalized so that the coefficient of $-\Box$ is unity.
The same holds for the ghosts and gravitino, so for each of these fields
the contributions of its spin components will have the same weight.
The bosonic ghosts contribute with a factor $-2$ relative to the graviton,
the gravitino with a factor $-1$, the fermionic ghost with a factor $2$ and the
Nielsen-Kallosh ghost with a factor $1$.
It remains to fix the identification of the spectral parameter with the cutoff.
As in section 3.2, for dimension two operators we will identify $t=k^{-2}$.
For dimension one and three operators, by the argument explained in section 3.3,
we will identify $u=2\pi^{-1/6} k^{-1}$ and $s=\frac{4}{3\sqrt{\pi}}k^{-3}$.
In this way, the beta function reads
\bea
\beta\!  &=&\! Y_{\Delta_{h^{TT}}}\!\left(\frac{1}{k^2}\right)\!
+Y_{\Delta_{\xi^T}}\!\left(\frac{1}{k^2}\right)\!
+Y_{\Delta_\sigma}\!\left(\frac{1}{k^2}\right)\!
+Y_{\Delta_h}\!\left(\frac{1}{k^2}\right)\!
-2Y_{\Delta_V}\!\left(\frac{1}{k^2}\right)\!
-2Y_{\Delta_S}\!\left(\frac{1}{k^2}\right)
\nn\\
&&
\!\!\!\!\!\!\!\!\!\!\!\!\!\!\!\!
-Y_{\Delta_{(\phi)}}\!\left(\frac{2}{\pi^{1/6}k}\right)\!
-Y_{\Delta{(\chi)}}\!\left(\frac{2}{\pi^{1/6}k}\right)\!
-Y_{\Delta_{(\psi)}}\!\left(\frac{2}{\pi^{1/6}k}\right)\!
+2Y_{\Delta_{(\eta)}}\!\left(\frac{2}{\pi^{1/6}k}\right)\!
+ Y_{\Delta_{NK}}\!\left(\frac{2}{\pi^{1/6}k}\right)\,.
\quad\quad\quad
\label{betasugra2}
\eea
We will now use this formula to obtain the beta function of pure supergravity.

Using \eq{euma}, \eq{beigen} and \eq{multiplicities}, the first few terms in the heat
kernel expansions for each bosonic spin operator are given by
\bea
\label{hkbos}
Y_{\Delta_{(h^{TT})}}(t)&=&\frac{V}{(4\pi t)^{3/2}}
\left(2-\frac{8}{3}R\,t+2\Lambda\,t\right)
+10+ \ldots
\nn\\
Y_{\Delta_{(\xi^{T})}}(t)&=&\frac{V}{(4\pi t)^{3/2}}
\left(2+\frac{2-3\alpha}{3}R\,t+6\alpha\Lambda\,t\right)
-5+ \ldots
\nn\\
Y_{\Delta_{(\sigma)}}(t)&=&\frac{V}{(4\pi t)^{3/2}}
\left(1+\frac{2}{3}R\,t+\frac{3\alpha}{4-\alpha}\Lambda\,t\right)
-5+ \ldots
\nn\\
Y_{\Delta_{(h)}}(t)&=&\frac{V}{(4\pi t)^{3/2}}
\left(1+\frac{1}{6}R\,t+\frac{12\Lambda}{4-\alpha}\right)
+ \ldots
\nn\\
Y_{\Delta_{(V)}}(t)&=&\frac{V}{(4\pi t)^{3/2}}
\left(2+\frac{2}{3}R\,t \right)
+ 1+ \ldots\ ,
\nn\\
Y_{\Delta_{(S)}}(t)&=&\frac{V}{(4\pi t)^{3/2}}
\left(1+\frac{16-\alpha}{6(4-\alpha)}R\,t\right)
-1+ \ldots
\eea

The terms $I_{-1}$ in (\ref{FI}) are all zero on account of the fact
that the coefficient of $n$ is twice the  coefficient of $n^2$ in all the sets of
eigenvalues. The ellipses stand for terms with positive powers of $t$.

Next, we list the results for the heat kernels for the fermions. They are
\bea
Y_{\Delta_{(\phi)}}(u)&=&
\frac{V}{\pi^2 u^3}\left[2- \left(\ft38 R - \ft14 \Lambda \right)u^2 \right]
+4+\ldots
\nn\\
Y_{\Delta_{(\chi)}}(u)&=&
\frac{V}{\pi^2 u^3}\left(2+\frac{(8-2\alpha'-\alpha'^2)R
-18\alpha'\sqrt{6\Lambda R}+54 \Lambda\alpha'^2}{24(1+\alpha')^2}\,u^2\right)-4+\ldots
\nn\\
Y_{\Delta_{(\psi)}}(u)&=&
\frac{V}{\pi^2 u^3}\left(2-\frac{(1+2\alpha'-8\alpha'^2)R
+18\alpha' \sqrt{6\Lambda R}-54\Lambda}{24(1+\alpha')^2}\,u^2\right)+\ldots
\nn\\
Y_{\Delta_{(\eta)}}(u)&=&
\frac{V}{\pi^2 u^3}\left(2+\frac{(8-2\alpha'-\alpha'^2)R
-18\alpha'\sqrt{6\Lambda R}+54 \Lambda\alpha'^2}{24(1+\alpha')^2}\,u^2\right)+\ldots
\nn\\
Y_{\Delta_{NK}}(u)&=&
\frac{V}{\pi^2 u^3}\left(2+\ft13 R\,u^2\right)
+\ldots
\label{hkfer}
\eea
Using the formula \eq{betasugra2} we obtain the total beta function
\bea
\beta_{SUGRA} &=& \frac{Vk^3}{(4\pi)^{3/2}} \Bigg[
\left(\frac{20+25\alpha-6\alpha^2}{4-\alpha}
-\frac{2}{\pi^{1/3}} \frac{5-4\alpha'}{1+\alpha'}\right)\widetilde \Lambda
\nn\\
&&
\qquad\qquad -\frac{1}{6}\!\left(\frac{92+7\alpha-6\alpha^2}{4-\alpha}
-\frac{2}{\pi^{1/3}}\,\frac{13+4\alpha'}{1+\alpha'}\right)\!\tilde R +O(\widetilde R^2)\Bigg]\ .
\label{betasugra}
\eea

A number of remarkable cancellations have occurred in obtaining \eq{betasugra}.
The $Y_0$ terms, which correspond to the integers outside the brackets
in \eq{hkbos} and \eq{hkfer}, cancel separately for the trace and tracefree parts
of $h_{\mu\nu}$, for the bosonic ghost, for the $\gamma$-trace and $\gamma$-tracefree
part of $\psi_\mu$ and for the fermionic and Nielsen-Kallosh ghosts.
Furthermore, the $Y_{-3/2}$ terms also cancel exactly, for the bosons and fermions
separately. This is related to the fact that the coefficient $Y_{-3/2}$ of each spin
is proportional to the number of corresponding degrees of freedom, and
in this theory there are no physical propagating degrees of freedom.
We shall discuss the consequences of these cancellations later on.


The expression (5.4) has a well-defined limit for $\alpha\to0$ and $\alpha'\to0$.
There is a subtlety if one tries to evaluate the beta function directly with $\alpha=0$
and $\alpha'=0$, because the unphysical fields $\xi_\mu^T$, $\sigma$ and $\chi$ are
not present in this gauge and the constant terms $+10$ and $+4$ in $Y_{\Delta_T}$
and $Y_{\Delta_\phi}$ seem to remain uncancelled.
In this gauge these terms are canceled in another way.
If one looks at the eigenvalues (4.14) in the gauge $\alpha=0$ one sees that
$\lambda_n^\xi=\lambda_n^V$ and $\lambda^\sigma_n=\lambda_n^S$, but in the spectrum of $V$ the
six zero modes with $n=1$ (i.e. the Killing vectors) are retained, while in the spectrum of $\xi$ they are absent.
As a consequence, $Y_{\Delta_\xi}=Y_{\Delta_V}-6$, and similarly
$Y_{\Delta_\sigma}=Y_{\Delta_S}-4$. Therefore, the bosonic contribution to (5.1) is
\be
Y_{\Delta_{(h^{TT})}}\!\left(\frac{1}{k^2}\right)\!
+Y_{\Delta_{(h)}}\!\left(\frac{1}{k^2}\right)\!
-Y_{\Delta_{(V)}}\!\left(\frac{1}{k^2}\right)\!
-Y_{\Delta_{(S)}}\!\left(\frac{1}{k^2}\right)-10\ .
\ee
The last term removes the constant term from the spin two sector.
In a similar way, in the fermionic sector $Y_{\Delta_\xi}=Y_{\Delta_\eta}-4$, where the four modes correspond to conformal Killing spinors, of which two are Killing spinors. So the fermionic contribution to (5.1) is
\be
-Y_{\Delta_{(\phi)}}\!\left(\frac{2}{\pi^{1/6}k}\right)\!
-Y_{\Delta_{(\psi)}}\!\left(\frac{2}{\pi^{1/6}k}\right)\!
+Y_{\Delta_{(\eta)}}\!\left(\frac{2}{\pi^{1/6}k}\right)\!
+ Y_{\Delta_{NK}}\!\left(\frac{2}{\pi^{1/6}k}\right)-4\,.
\ee
The result is the same as taking the limit in (5.4).

We note that the effective action in physical gauge can be derived directly from a change of variables in the functional integral, bypassing the standard Faddeev--Popov construction \cite{Bern}. This procedure has recently been applied to three--dimensional gravity in \cite{Gaberdiel:2010xv} by using the results given in sections 4.1.2 and 4.2.2.

As discussed in section 3.3, by comparing \eq{betasugra} with \eq{mr} we read off
the coefficients $A$, $B$ and $C$ as follows:
\bea
A &=& \frac{2}{\sqrt\pi} \left(\frac{20+25\alpha-6\alpha^2}{4-\alpha}
-\frac{2}{\pi^{1/3}} \frac{5-4\alpha'}{1+\alpha'}\right)\widetilde \Lambda\ ,
\label{A}\w2
B &=& -\frac{1}{3\sqrt\pi} \left(\frac{92+7\alpha-6\alpha^2}{4-\alpha}
-\frac{2}{\pi^{1/3}}\,\frac{13+4\alpha'}{1+\alpha'}\right)\ ,
\label{B}\w2
C&=&0\ .
\eea
The beta functions of $\widetilde\Lambda$, $\widetilde G$ and $\tilde \mu$
are given in \eq{oneloopbeta} with the above values $A$, $B$ and $C$.
The vanishing of $C$ follows from the  cancellation of the $Y_0$ terms. It implies that a Chern-Simons term
is not generated by quantum corrections at one loop. Due to the cancellation of the leading terms $Y_{-3/2}$,
the beta function of $\widetilde\Lambda$ is proportional to $\Lambda$ itself.
As a result, $\widetilde\Lambda=0$ is a fixed point.
This is the same as in bosonic three-dimensional Einstein gravity \cite{Percacci:2010yk}.
The existence of a fixed point in the beta function for $\widetilde G$ requires $B$ to be negative.
This is true for any value of $\alpha$ and $\alpha'$ in the ranges specified earlier in \eq{alpharange} and \eq{betarange}. Then the fixed point is at $\widetilde G=-1/B$.
In the gauge $\alpha=0$, $\alpha'=0$ the numerical position of the fixed point is
$(\widetilde\Lambda,\widetilde G)=(0,1.013)$; it is attractive in both directions,
with scaling exponents $-1$ in the $\widetilde G$ direction and $-4.045$ in the $\widetilde\Lambda$ direction.
The flow of pure supergravity is depicted in figure 1, left panel.

We also observe that on shell, i.e. for $R=6\Lambda$, the whole beta function becomes independent of the gauge parameters $\alpha$ and $\alpha'$, as expected:
\be
\beta_{SUGRA}^{on-shell} \!=\! \frac{2Vk^3}{(4\pi)^{3/2}}
\left(\frac{8}{\pi^{1/3}}-9\right) \widetilde\Lambda\ .
\ee
%

%
%
%

\section{The Beta Functions of Topologically Massive \hfill\break Supergravity }


From \eq{bops} and \eq{fops} we see that the Chern-Simons term and its superpartner contribute to the
wave operators of the spin-2 and spin-3/2 fields only. The computations for the lower-spin
sectors of the preceding section will not be affected. Therefore, in this section we will focus on the heat kernels of the spin-2 and spin-3/2 operators in the presence of the Chern-Simons term proportional to $\mu^{-1}$. The eigenvalues for the spin-2 field are now third-order polynomials in $n$, and the Euler-Maclaurin integrals of the form $\int dx d_x e^{-\lambda_x}$ (see \eq{hk} and \eq{euma}) cannot be computed in closed form. In what follows, we shall compute these integrals for the two cases of
large $\tilde\mu$, and small $\tilde\mu$, separately, where $\tilde\mu=\mu/k$.


\subsection{The Large $\tilde\mu$ Limit}


In this limit, we can treat the contribution of the Chern-Simons term as a small perturbation of the results
for pure supergravity discussed above.
For the bosons, using the eigenvalues and multiplicities in \eq{beigen}, \eq{multiplicities},
the integral term in \eq{euma} can be expanded in $\rho/\mu$,
yielding for the two polarization states
\be
\int_2^\infty\!\!\! dx (x^2+2x-3)
e^{-t(\rho^2(x^2+2x+1)-\Lambda)} \!
\left[1\mp\frac{t \rho^3}{\mu}x(x+1)(x+2)
+\ldots\right]\ .
\ee
Summing the contributions of the positive and negative spin-2 polarizations, including also the Bernoulli sums in
\eq{euma}, the odd powers of $1/\mu$ cancel and to order $1/\mu^2$ this leads to the result
\be
\label{james}
Y_{\Delta_{(h^{TT})}}(t) = \frac{V}{(4\pi t)^{3/2}}
\left( 2 -\frac{8}{3}R\,t +2\Lambda\,t+ \frac{105}{8\mu^2 t} -\frac{15 R}{4\mu^2}
+\frac{105 \Lambda}{8\mu^2}
\right) + 10+\ldots
\ee
The ellipses refer to terms that contain increasing powers of $t$ but also $1/t$, the latter coming from the increasing powers of $x$ in the integral. One should obviously not regard this as an expansion for arbitrarily small $t$, rather, the expansion is valid for $\frac{1}{\mu^2}\ll t\ll\frac{1}{R}$.

Similarly, for the fermions, using the eigenvalues and multiplicities
given in \eq{feigen} and \eq{fmult}, and expanding the integrand occurring in \eq{euma}
in $\rho/\mu$, we get
\be
\int_{0}^\infty dx\, (x+1)(x+4)e^{-u[\rho (x+\frac52)\mp\frac{1}{2} m]}\
\left[1 \mp\frac{u \rho^2}{\mu}(x+2)(x+3)
+\ldots\right]\ .
\label{fred}
\ee
Note that convergence for positive $u$ requires that the eigenvalues
should tend to $+\infty$ for large $n$.  In the case of the eigenvalues
$\lambda_n^{(\phi)-}$, which tend to $-\infty$, we have reversed their overall sign.
(Since we are interested in the scaling behaviour of the (regularised)
determinant $\prod_n\lambda_n$, an overall sign reversal of the $\lambda_n$
has no material effect.) This leads to the result
\bea
\label{jane}
Y_{\Delta_{(\phi)}} (u) &=&
\frac{V}{\pi^2 u^3}\Bigg[ 2- \left(\frac{3}{8} R
- \frac{1}{4} \Lambda \right)u^2
+\frac{\sqrt{\Lambda}}{\mu} \left(-12+\frac{5}{12} R u^2 -\frac{1}{2} \Lambda u^2 \right)
\nn\\
&&
+ \frac{1}{\mu^2}\left( -\frac{11}{2} R
+45\Lambda +\frac{360}{u^2}
\right)\Bigg] +4 +\ldots
\eea
Thus, the $\mu$-dependent contribution to the beta function, to order $1/\tilde\mu$, is
\be
\Delta\beta_\mu = \frac{Vk^3}{(4\pi )^{3/2}} \Bigg[
\frac{4\sqrt{\widetilde\Lambda}}{\pi^{1/3}\tilde\mu}
\left(3\pi^{1/3}+\ft{1}{2} \widetilde\Lambda-\ft{5}{12}\widetilde R  \right)
+O\left(\frac{1}{\tilde\mu^2}\right)
\Bigg]\ .
\ee
Notice that the inverse powers of $t$ and $u$ in \eq{james} and \eq{jane} have become positive powers of $k$ which combine with powers of $1/\mu$ to produce an expansion in $1/\tilde\mu$.
The total beta function for topologically massive supergravity in the large $\tilde\mu$ limit is
\bea
\beta_{TMSG} &=& \beta_{SUGRA}+\Delta\beta_\mu
\nn\\
&=& \frac{Vk^3}{(4\pi)^{3/2}} \Bigg\{
\left(\frac{(20+25\alpha-6\alpha^2)}{4-\alpha}
-\frac{2}{\pi^{1/3}} \frac{5-4\alpha'}{1+\alpha'}\right)\widetilde \Lambda
+\frac{1}{\tilde\mu}\left(12{\widetilde\Lambda}^{1/2}
+\frac{2}{\pi^{1/3}}\, {\widetilde\Lambda}^{3/2}\right)
\nn
\\
&&
\!\!\!\!\!\!\!\!\!\!\!\!
+\left[ -\frac{92+7\alpha-6\alpha^2}{6(4-\alpha)}
+\frac{1}{3\pi^{1/3}} \frac{13+4\alpha'}{1+\alpha'}
-\frac{5}{3\pi^{1/3}} \frac{\widetilde\Lambda^{1/2}}{\tilde\mu}
\right]\widetilde R
\Bigg\}\ .
\label{betalarge}
\eea

For ${\widetilde\mu} \to \infty$, this agrees with \eq{betasugra}. Regarding on-shell gauge-parameter independence, we observe that this had already been shown for $\beta_{SUGRA}$ and that the correction terms $\Delta\beta_\mu$, are gauge parameter-independent even off-shell, since they derive entirely from
the spin-2 and spin-3/2 contributions.


\subsection{The Small $\tilde\mu$ Limit}


In the regime where $\mu$ is small relative to $k$,
the cubic term in the spin-2 wave operator is dominant and we can consider the quadratic term as a small perturbation.  Likewise, for the spin-3/2 operator the quadratic term is dominant. We therefore replace the operators $\Delta_{(h^{TT})}$ and $\Delta_{(\phi)}$ by $\mu\Delta_{(h^{TT})}$ and $\mu\Delta_{(\phi)}$ respectively, so that the leading-order terms have dimensionless coefficients. Correspondingly, we use the spectral parameter $s$, which has dimension $L^3$, for the spin-2 operator, and $t$, with dimension $L^2$, for the spin-3/2 operator. Thus, to evaluate the heat kernel for spin-2, in the integral in \eq{euma} we expand the eigenvalues in the exponential and obtain
\be
\int_2^\infty\!\!\! dx (x^2+2x-3)
e^{-s\rho^3x(x+1)(x+2)} \!
\left[1\mp s\mu(\rho^2(x^2+2x+1)-\Lambda)
+\ldots\right]\,.
\ee
Note that, following the same logic as in \eq{fred}, for convergence we have changed the
overall sign of the eigenvalue when summing over $\lambda_n^{h^{TT}-}$.
Summing the contributions of the positive and negative spin-2 polarizations
the odd powers of $\mu$ cancel and keeping terms up to order $\mu^2$ one obtains
\footnote{In practice the integral with a cubic polynomial in the exponent
is still too hard. We get around this difficulty by keeping only the
cubic term in the exponential and Taylor expanding the exponential of the
quadratic and linear term.}
\be
Y_{\Delta_{(h^{TT})}}(s)=
\frac{V}{6\pi^2 s}
\left(2+ \frac13\Gamma(\ft43)\left(4\mu^2-11R\right)\,s^{2/3}\right)+10+\ldots
\label{sum1}
\ee
Similarly for the fermions, using the eigenvalues in \eq{feigen} and expanding the integrand occurring in \eq{euma}, we obtain
\be
\int_{0}^\infty dx\, (x+1)(x+4)e^{-t\rho^2(x+2)(x+3)}\
\left[1 -t\mu\left(\pm\rho\left(x+\ft52\right)-\ft12 m\right)
+\ldots\right]\ .
\ee
The sum of the positive and negative spin 3/2 polarizations gives
\be
Y_{\Delta_{(\phi)}}(t) =
\frac{V}{(4\pi t)^{3/2}}\left(2 +\frac{1}{12}\left(-17R +6\mu(2\sqrt\Lambda +3\mu)\right) t
+\ldots\right) +4\ .
\label{sum2}
\ee
With the use of $s$ and $t$, as opposed to $t$ and $u$, as the spectral parameters, the formula \eq{betasugra2} for the total beta function is now replaced by
\bea
\beta\!  &=&\! Y_{\Delta_{h^{TT}}}\!\left(\frac{4}{3\sqrt{\pi}k^3}\right)\!
+Y_{\Delta_{\xi^T}}\!\left(\frac{1}{k^2}\right)\!
+Y_{\Delta_\sigma}\!\left(\frac{1}{k^2}\right)\!
+Y_{\Delta_h}\!\left(\frac{1}{k^2}\right)\!
-2Y_{\Delta_V}\!\left(\frac{1}{k^2}\right)\!
-2Y_{\Delta_S}\!\left(\frac{1}{k^2}\right)
\nonumber
\\
&&
\!\!\!\!\!\!\!\!\!\!\!\!\!\!\!\!\!\!\!\!\!\!
-Y_{\Delta_{(\phi)}}\!\left(\frac{1}{k^2}\right)\!
-Y_{\Delta{(\chi)}}\!\left(\frac{2}{\pi^{1/6}k}\right)\!
-Y_{\Delta_{(\psi)}}\!\left(\frac{2}{\pi^{1/6}k}\right)\!
+2Y_{\Delta_{(\eta)}}\!\left(\frac{2}{\pi^{1/6}k}\right)\!
+ Y_{\Delta_{NK}}\!\left(\frac{2}{\pi^{1/6}k}\right)\,.
\label{betasugra3}
\eea
Putting together the above results in the formula \eq{betasugra3}, we obtain the total beta function
\bea
\beta_{TMSG}\!\! &=&\!\! \frac{Vk^3}{(4\pi)^{3/2}} \Bigg[
- \tilde\mu \sqrt{\widetilde\Lambda}
+ \left(\frac{3(4+9\alpha-2\alpha^2)}{4-\alpha}
-\frac{9}{\pi^{1/3}} \frac{1-\alpha'}{1+\alpha'}  \right)      \widetilde \Lambda
\\
&&
\qquad\qquad
+\left( \frac{4-21\alpha+4\alpha^2}{4(4-\alpha)}
+\frac{1}{6\pi^{1/3}} \frac{1-17\alpha'}{1+\alpha'} -\frac{44\,\Gamma\left(\ft43\right)}{3\pi^{1/3}6^{2/3}}
\right)\widetilde R
+O(\tilde\mu^2)
\Bigg]
\nn
\label{betasmall}
\eea
Note that the limit $\tilde\mu\rightarrow 0$ can be taken without difficulty. We observe that the leading, curvature-independent, term is no longer proportional to $\widetilde\Lambda$.
On-shell, the beta function is again gauge-parameter independent:
\bea
\beta_{TMSG}^{on\,shell}\!\! &=&\!\! \frac{Vk^3}{(4\pi)^{3/2}} \Bigg[
- \tilde\mu \sqrt{\widetilde\Lambda}
+ \left(\frac92 -\frac{88\,\Gamma(\ft43)}{\pi^{1/3}6^{2/3}}  +\frac{8}{\pi^{1/3}}\right)
 \widetilde \Lambda+O(\tilde\mu^2)
\Bigg]\ .
\eea


\section{The RG Flows}


Comparing the results \eq{betalarge} and \eq{betasmall}
with \eq{mr}, one can read off the coefficients $A$, $B$ and $C$,
and write out the beta functions as in \eq{oneloopbeta}.
It turns out that $A$ and $B$ are functions of $\tilde\mu$ and $\widetilde\Lambda$.
Due to the cancellation of the $Y_0$ terms (separately for bosons and fermions),
the coefficient $C$ is zero. This result is independent of the shape of
the cutoff and is therefore a truly universal feature of the theory.
It implies that the dimensionless combination
\be
\nu \equiv\mu G
\ee
has vanishing beta function. Since $\nu$ does not run, in equations \eq{oneloopbeta} we can replace
$\tilde\mu$ by $\nu/\widetilde G$, with $\nu$ constant.
The beta functions for ${\widetilde G}$ and ${\widetilde\Lambda}$ thus have the form
\bea
k\, \fft{d\tilde G}{dk} &=&
\widetilde G+B(\widetilde\Lambda,\nu/\widetilde G) \widetilde G^2\ ,
\nonumber
\\
k\, \fft{d\tilde\Lambda}{dk} &=& -2\tilde\Lambda
+\frac{1}{2}A(\widetilde\Lambda,\nu/\widetilde G)\widetilde G+B(\widetilde\Lambda,\nu/\widetilde G) \widetilde G\tilde\Lambda\ .
\label{betas}
\eea
This system describes a flow in the $\widetilde\Lambda$-$\widetilde G$ plane, depending on the fixed external parameter $\nu$, as well as the gauge parameters $\alpha$ and $\alpha'$.
We shall now analyse these flows in the large and small $\nu$ approximations, using the beta functions presented above.


\subsection{The Large $\nu$ Limit}


%
\begin{figure}
\begin{center}
{
\resizebox{0.5\columnwidth}{!}
{\includegraphics{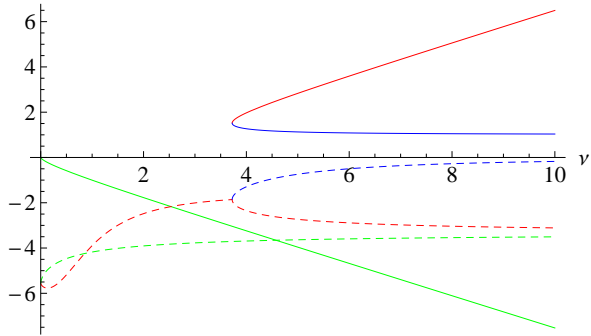}}
\caption{\small Position of the fixed points in the large $\nu$ approximation.
The red, green, blue continuous curves give the value of $\widetilde G_*$
for the three solutions, the dashed curves give the corresponding
values of $\widetilde\Lambda_*$. Only the blue solution is reliable,
the remaining two are artifacts of the approximation.}
}
\end{center}
\end{figure}

Since $\nu=\tilde\mu\widetilde G$, for any fixed finite $\widetilde G$
the large $\tilde \mu$ expansion is also a large $\nu$ expansion.
Conversely, for $\nu\gg1$ we can use the results of subsection 6.1 to gain information
on the flow in the $\widetilde\Lambda$-$\widetilde G$ plane for $\tilde G$ of order one
or smaller.
From equations \eq{mr} and \eq{betalarge} we read off
\bea
A &=& \frac{2}{\sqrt\pi} \left(\frac{20+25\alpha-6\alpha^2}{4-\alpha}
-\frac{2}{\pi^{1/3}} \frac{5-4\alpha'}{1+\alpha'}\right)\widetilde \Lambda
+\frac{4}{\pi^{5/6}\nu}\widetilde G \sqrt{\widetilde\Lambda}(\widetilde\Lambda +6\pi^{1/3})\ ,
\label{AA}\\
B &=& -\frac{1}{3\sqrt\pi} \left(\frac{92+7\alpha-6\alpha^2}{4-\alpha}
-\frac{2}{\pi^{1/3}} \frac{13+4\alpha'}{1+\alpha'}\right) +\frac{20}{3\pi\nu}\widetilde G \sqrt{\widetilde\Lambda} \ ,
\label{BB}\\
C&=&0\ ,
\label{rgflows}
\eea
\goodbreak
where we have kept only the leading term in $1/\nu$.
Up to this order we see that the coefficient $A$
vanishes for $\widetilde\Lambda=0$, so any fixed point
in the $\widetilde\Lambda$-$\widetilde G$ plane will be at $\widetilde\Lambda=0$. From \eq{betalarge}, we see, however, that at order $1/\nu^2$ this property generically does not hold.

For $\nu\to\infty$, the results go over to those of pure supergravity with cosmological constant which we discussed in section 5.2. A new feature that arises for finite but nonvanishing values of $\nu$ is that the flow equation for $\widetilde G$ now depends on $\widetilde\Lambda$.
For $\nu^{-1} \not=0$, the fixed point of pure supergravity gets shifted by
a small amount in the negative $\widetilde\Lambda$ direction.

Since $A$ and $B$ contain terms proportional to $\widetilde G$, the fixed point equations
are cubic (see \eq{rgflows}) and will generically admit three solutions.
The position of these solutions is plotted in figure 1.
The continuous and dashed blue curves give the values of $\widetilde\Lambda$
and $\widetilde G$ for the solution that asymptotes to the SUGRA solution.
For $\nu>3.7$ the additional two solutions are real, one (red) with positive,
and one (green) with negative $\widetilde G$.
For these solutions $|\widetilde G|$ grows linearly
with $\nu$ with a coefficient of order one, therefore $\tilde\mu\approx 1$ and
they occur outside the domain where the approximation is reliable.
For $\nu\approx 3.7$ one of these solutions merges with the one that asymptotes
to SUGRA, and they become complex, but at this low value of $\nu$
the approximation is unreliable even for $\widetilde G$ of order one.
A picture of the flow for $\nu\to\infty$ (pure supergravity) and $\nu=10$,
in the region of the plane where the approximation is reliable, is shown in figure 2.

%
\begin{figure}
\begin{center}
{\resizebox{0.66\columnwidth}{!}
{\includegraphics{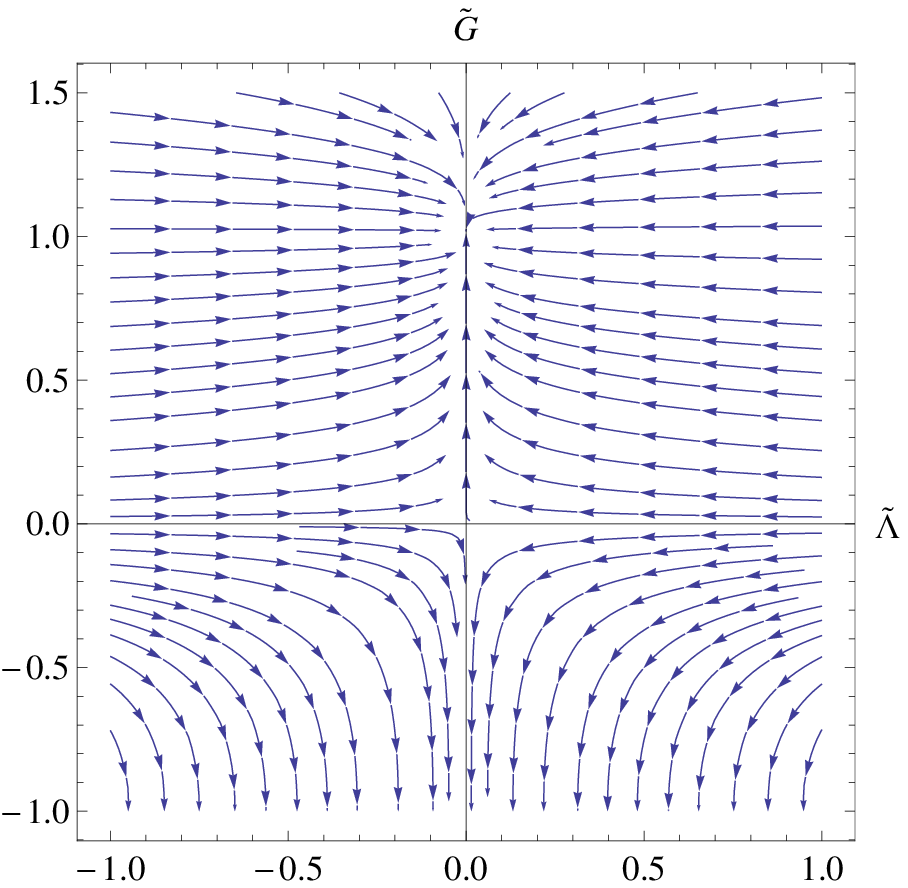}
\includegraphics{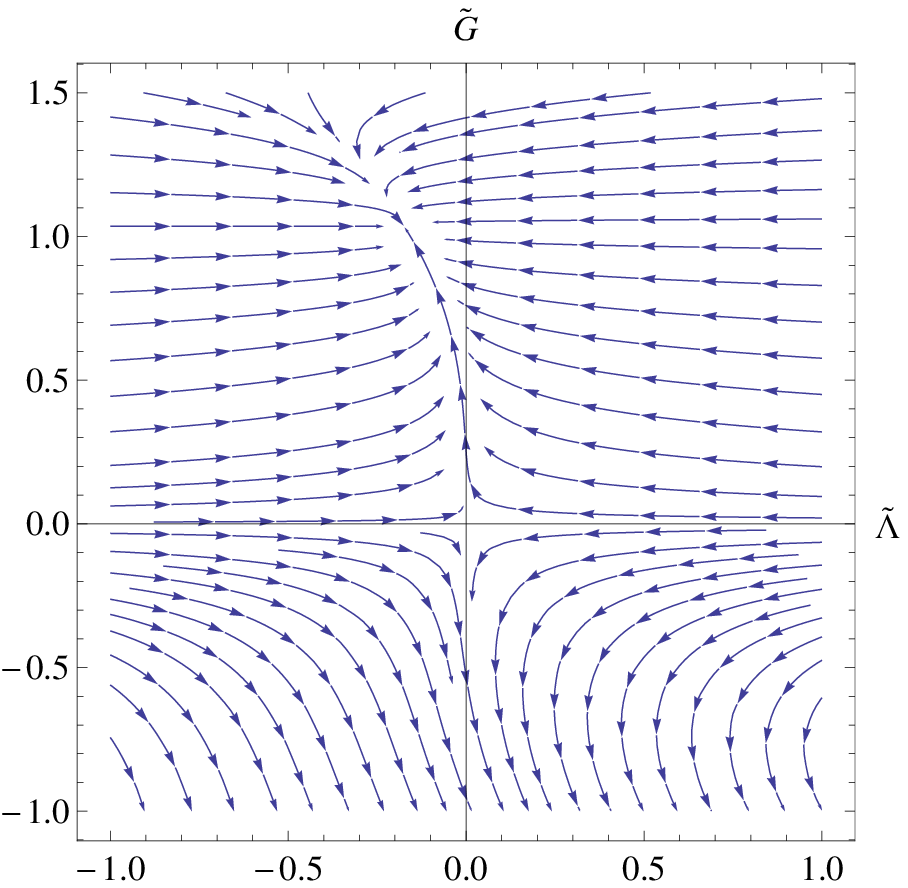}}
\caption{\small
Flows in the $\widetilde\Lambda$-$\widetilde G$ plane
in the gauge $\alpha=0$, $\alpha'=0$, and large $\nu$.
Left: pure SUGRA ($\nu\to\infty$); right: $\nu=10$.}}
\end{center}
\end{figure}
%
%

\subsection{The Small $\nu$ Limit}


Since $\nu=\tilde\mu\widetilde G$, for finite $\widetilde G$
the small $\tilde \mu$ expansion is also a small $\nu$ expansion.
Conversely, for $\nu\ll 1$ we can use the results of subsection 6.1 to gain information
on the flow in the $\widetilde\Lambda$-$\widetilde G$ plane for $\widetilde G$ of order one
or larger.
From equations \eq{mr} and \eq{betasmall} we read off
\bea
A &=& \frac{2}{\sqrt\pi} \left[
\frac{3(4+9\alpha-2\alpha^2)}{4-\alpha}
-\frac{9}{\pi^{1/3}} \frac{1-\alpha'}{1+\alpha'} \right]\widetilde\Lambda
-\frac{2\nu}{\sqrt\pi} \frac{\sqrt{\widetilde\Lambda}}{\widetilde G}\ ,
\nn\\
B&=& \frac{2}{\sqrt\pi}\Bigg[\frac{4-21\alpha +4\alpha^2}{4(4-\alpha)}
+\frac{1}{6\pi^{1/3}}\frac{1-17\alpha'}{1+\alpha'} -\frac{44\,\Gamma\left(\ft43\right)}{3\pi^{1/3}6^{2/3}} \Bigg]\ ,
\nn\\
C&=&0\ .
\eea
We have kept only the leading term in $\nu$. Even though there is just one term arising in $A$ that depends on $\nu$, it should be stressed that the $\nu$-independent parts are not those of pure supergravity with cosmological constant, and their form depend on the Chern-Simons term.

The limit $\nu\to 0$ can be taken without difficulty and results in a flow with two fixed points:
the usual Gaussian fixed point and a non-Gaussian one.
In any gauge, the Gaussian fixed point, which is at the origin, has scaling exponents equal to the canonical dimensions:
$1$ in the $\widetilde G$ direction and $-1$ in the $\widetilde\Lambda$ direction.
In the gauge $\alpha=0$ and $\alpha'=0$ the non-Gaussian fixed point occurs
at $\widetilde\Lambda=0$, $\widetilde G=1.692$
and it has scaling exponents $-1$ in the $\widetilde G$ direction
and $-6.003$ in the $\widetilde\Lambda$ direction.

For $\nu\not=0$ the flow develops a singularity for $\widetilde G\to 0$
and the Gaussian fixed point seems to disappear,
but we recall that the picture of the flow is not reliable in this limit.
A picture of the flow for $\nu=0$ and for $\nu=0.1$ is given in figure 2.

%
\begin{figure}
\begin{center}
{\resizebox{1\columnwidth}{!}
{\includegraphics{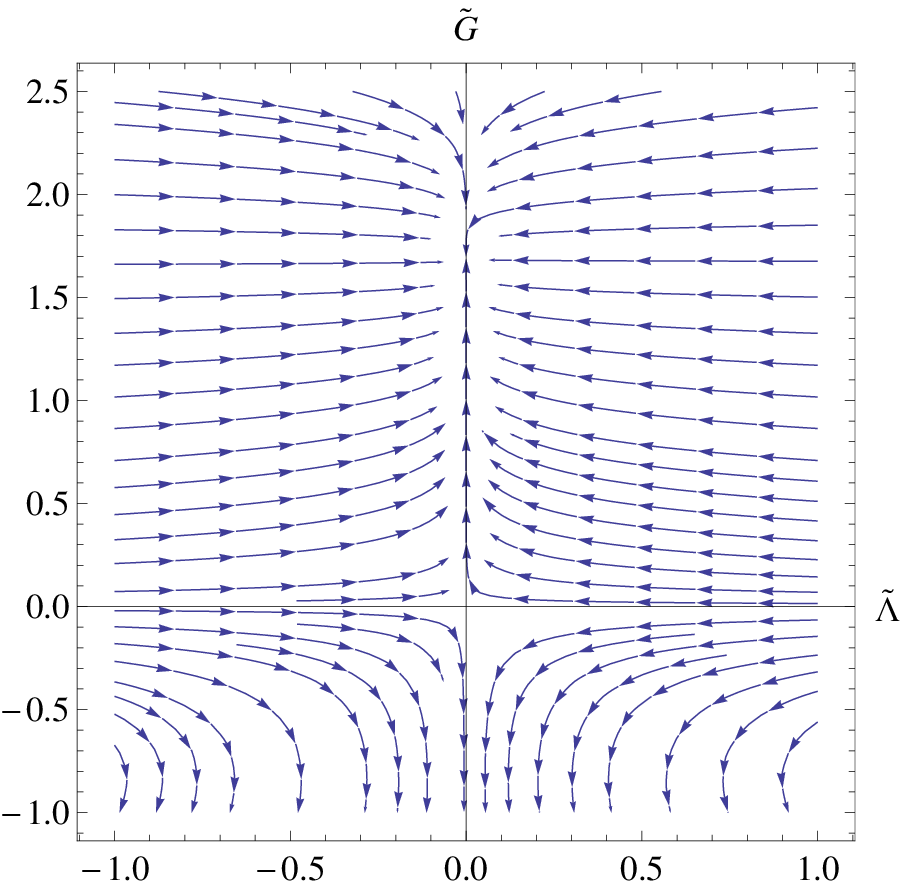}
\includegraphics{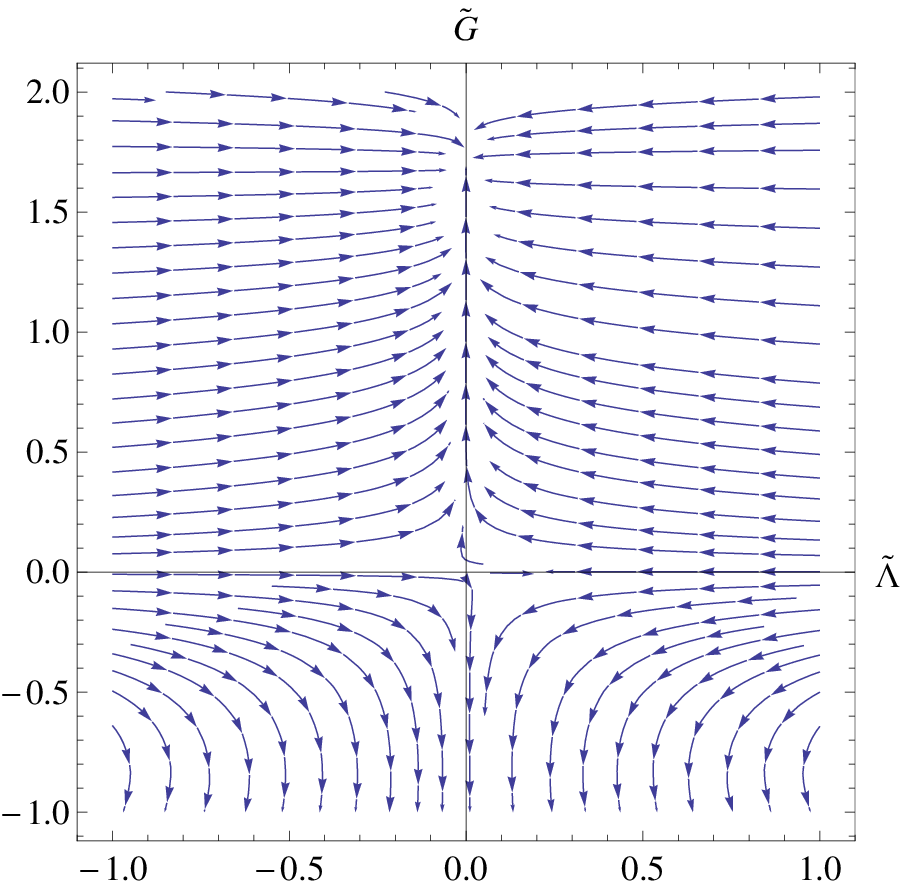}
\includegraphics{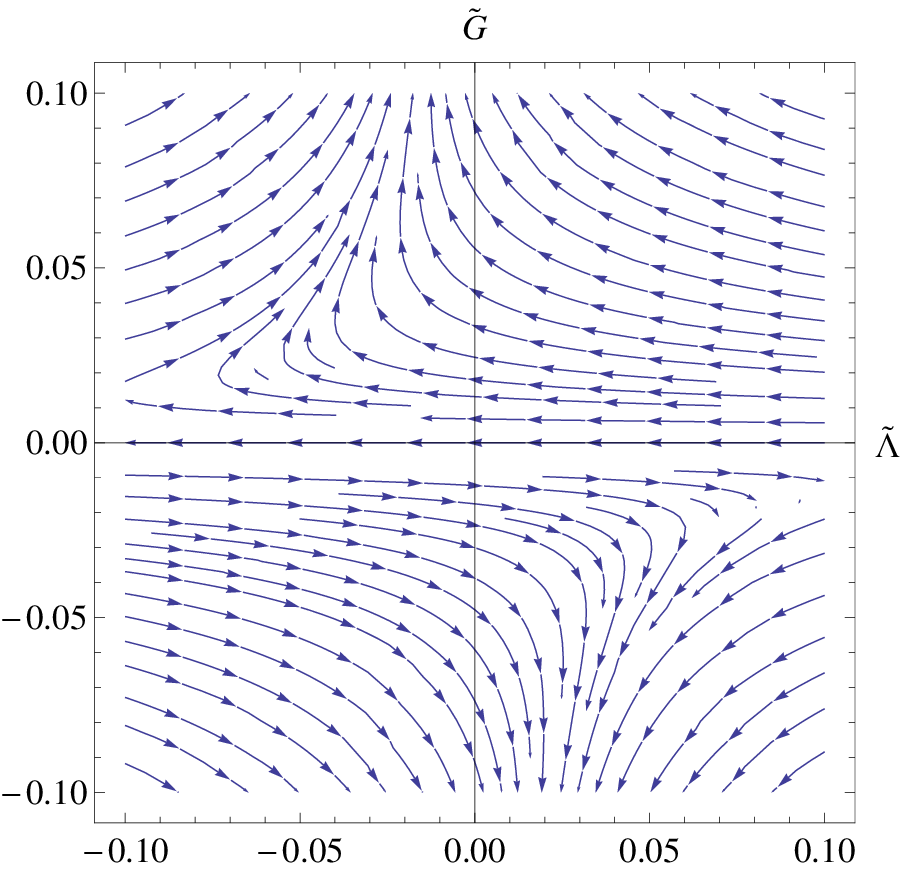}
}
\caption{\small Flows in the $\widetilde\Lambda$-$\widetilde G$ plane
in the gauge $\alpha=0$, $\alpha'=0$, for $\nu=0$ (left) and $\nu=0.1$ (middle).
The shift of the nontrivial fixed point is too small to be seen on this scale,
but one notices a different behavior near the origin.
An enlargement of this area (right panel) reveals that the Gaussian fixed point
is absent for $\nu=0.1$. In fact, the beta functions become singular on the
$\widetilde\Lambda$ axis. This, however, is an artifact of the approximation,
which breaks down when $\widetilde G$ becomes too small, in this case of order $0.01$.
}}
\end{center}
\end{figure}
%
%


\section{Conclusions}


We have calculated the renormalization group beta functions for topologically massive supergravity
in three dimensions. Logarithmic divergences in four dimensional 
supergravities
have been computed previously using heat kernel methods for example in \cite{rocek,Fradkin:1981iu,macarthur}.
However, these calculations were limited to second order wave operators of Laplace type.
Here we have been able to deal with a third order operator for which the heat kernel
coefficients are not available and in addition we have calculated also the power law divergences.
We have found that, as expected, $\nu$, the coefficient of the Chern-Simons term, does not get renormalized.
This accords with the notion that the coefficient of the Chern--Simons term is quantized,
at least for suitable boundary conditions.
\footnote{It has been argued in \cite{Percacci:1986hu} that $\nu$ need not be quantized
on the three-sphere, but unless the topology is fixed a priori, it is enough to
find one topology where large gauge transformations exist to impose quantization of $\nu$.}
The flow of the cosmological constant and Newton's constant depends parametrically on $\nu$.
We have studied their behavior in the limit of $\nu$ very large or very small.
The qualitative picture of the flow is similar to that encountered in TMG \cite{Percacci:2010yk},
having both a Gaussian and a non-Gaussian fixed point. The latter occurs for vanishing cosmological constant
and positive Newton's constant in both limits.

We now return to the question raised in the introduction, namely whether the generic theory
flows to the chiral point.
With the quantization procedure described here, which makes sense for generic values
of the couplings, we find that the ratio $\mu/\sqrt{\Lambda}=1$ is not preserved by the flow.
It would be interesting to quantize the chiral TMG (or TMSG) and to determine whether
its RG flow preserves the chirality condition.

It has been argued \cite{oda,oda2,oda3} that TMG is renormalizable.
In this case there must be a neighborhood of the origin in the $\Lambda$-$G$ plane
where the picture of the flow given in figure 2 is correct to all orders.
However, perturbative renormalizability is not sufficient to guarantee the existence of the theory:
only asymptotically free theories can be proven to exist by perturbative methods.
In the present case, a glance at figure 2 shows that in the neighborhood of the Gaussian
fixed point, the $G$-direction is not asymptotically free.
Thus, if one starts anywhere with $\widetilde G>0$, $\widetilde G$ will grow.
The question is whether this growth leads the theory outside the domain of perturbation theory or not.
Our calculations seem to imply that the theory tends to a non-Gaussian fixed point,
and that the growth of $\widetilde G$ ceases. 

   It would be interesting to extend our results to other three-dimensional
models that
contain higher-order curvatures \cite{ohta}, as well as the 
conformally-invariant
model discussed in \cite{grum1,grum2}, where only the Chern-Simons term
survives.  In this latter case one cannot simply take the $\mu\rightarrow 0$
limit of our results for the beta function, since the additional local
Weyl symmetry would have to be built into the quantisation procedure
from the outset.  

There are a number of issues related to background supersymmetry and various scheme dependences in the calculation of the beta functions. Firstly, in the off-shell computations, by which we mean those in which the on-shell equation $R=6\Lambda$ is not used, the total quadratic action including the gauge fixing and ghost actions is clearly not invariant under the {\it rigid} background supersymmetry transformations \eq{bsusy}. In view of the results of appendix F, this symmetry cannot be present on-shell either. This is not a problem, however, since the symmetry in question is a rigid one. This state of affairs arises in all quantum supergravity computations performed in their component formulations; see, for example \cite{Fradkin:1983mq}. The study of this issue by means of the background field method in curved superspace, and in backgrounds that are not purely bosonic, is beyond the scope of this paper.

Secondly, the beta functions of $G$ and $\Lambda$ depend on the choice of cutoff profile function, which we have chosen to be simply a Heaviside theta function. To compare with approaches adopted in the literature, we note that in the context of beta functions in $N=4, D=4$ gauged supergravity, $\zeta$-function regularization has been used in \cite{Fradkin:1983mq,Granda:1997xk}, and the method of modifying the kinetic term by the introduction of a suitable term in the total quadratic action has been used in \cite{Bytsenko:1994at}. As is well known in the case of $\zeta$-function regularization, only the logarithmic divergences can be probed, and it is not useful for the computation of the beta functions for dimensionful couplings.

Thirdly, there is a choice to be made in relating the spectral parameters $s$, $t$ and $u$ \eq{stu} to the cutoff $k$.
We have made a natural choice such that the contributions to the beta function of the cosmological
constant (which is proportional to the leading terms of the heat kernels) are proportional to the
number of degrees of freedom of each field.


Finally,  dependence of the beta functions on the choice of gauge parameters in the gauge fixed action
is to be expected off-shell, but we have shown that there is no dependence on shell. We refer the reader to ref. \cite{Fradkin:1983mq} for a discussion of various issues that arise in the context of the expected gauge dependence.

Of course, all these arbitrary choices must become immaterial when one calculates
a physical observable, but the quantities whose beta functions have been calculated here
are not sufficient.
The reason is simply that in the calculation of a typical field theoretic observable,
even to lowest order in perturbation theory, there are contributions from
terms in the effective action that are not accounted for here.
For example, to calculate a typical $n$-point function at one loop it is enough to know the effective action to $n$-th order in the field, but one has to retain the full momentum dependence. By contrast, in our calculation we are retaining the full field dependence but we truncate the momentum dependence to the third order. For a discussion of the difficulties that arise when one tries to convert the cutoff dependence of couplings into momentum dependence of physical observables, see \cite{anber}.

\subsection*{Acknowledgment}

E.S. thanks SISSA, and R.P. thanks Texas A\&M University for hospitality 
during the preparation of this work.  M.J.P. would like to thank the 
Mitchell foundation and Trinity College Cambridge for their generous support. 
The research of E.S. is supported in part by NSF grants PHY-0906222 and 
PHY-1214344. C.N.P. is supported in part by DOE grant DE-FG03-95ER40917. 
M.J.P. is in part supported by the STFC rolling grant STJ000434/1.

\bigskip

\goodbreak

\begin{appendix}

\section{Variational Formulae}

The first variation of the Einstein-Hilbert action in $d$-dimensional spacetime, up to total derivative terms assumed to integrate to zero, is given by
\bea
\delta \int d^d x \sqrt{-g} R &=& \int d^d x \sqrt{-g}\left(-R^{\mu\nu}+\frac12 g^{\mu\nu} R+\nabla^\mu\nabla^\nu +g^{\mu\nu} \Box\right)\delta g_{\mu\nu}
\nn\\
&=& \int d^d x \sqrt{-g}\left(-2R^{\mu\nu}+g^{\mu\nu} R\right) e_\mu{}^a\delta e_{\nu a}\ .
\eea
 The second variation, for an arbitrary background and up to total derivative terms, assumed to integrate to zero, and using the notation $\delta g_{\mu\nu}=h_{\mu\nu}$, yields
\bea
\delta^2 \int d^d x \sqrt{-g} R &=& \int d^d x \sqrt{-g}\Bigg[
\left(-2R^{\mu\nu}+g^{\mu\nu} R\right) \delta e_\mu{}^a\delta e_{\nu a}
\nn\\
&& -\frac12 h^{\mu\nu}\nabla_L h_{\mu\nu} +(\nabla^\sigma h_{\mu\sigma})^2 +h\nabla^\mu \nabla^\nu h_{\mu\nu}
-\frac12 h\Box h \Bigg]
\nn\\
&& -hR^{\mu\nu}h_{\mu\nu} -\frac12 Rh^{\mu\nu}h_{\mu\nu} +\frac14 Rh^2 +2R_{\mu\nu} h^{\mu\alpha} h^\nu{}_\alpha\Bigg]\ ,
\eea
where $h_{\mu\nu} =2e_{(\mu}{}^a\delta e_{\nu) a}$ and $h^{\mu\nu}= g^{\mu\rho} g^{\nu\sigma}h_{\rho\sigma}$ and
\be
\nabla_L h_{\mu\nu}=-\Box h_{\mu\nu} -2R_{\mu\rho\nu\sigma} h^{\rho\sigma} +R_{\mu\rho} h^\rho{}_\nu
+ R_{\nu\rho} h^\rho{}_\mu\ .
\ee
%

\section{Exponential Cutoff}

   An alternative choice is to use a smooth cutoff rather than a
step function.  A natural possibility that one might consider
is the exponential function, $\tilde C(\tilde t) =e^{-\tilde t}$,
since this indeed tends rapidly to zero at large $\tilde t$, and
it approaches 1 as $\tilde t$ tends to zero.  Unfortunately
$e^{-\tilde t}$ does not approach 1 sufficiently rapidly at small $\tilde t$.
For our present purposes, it
turns out that $\tilde C(\tilde t) =e^{{-\tilde t}^2}$ will work.  In
order to encompass more general situations, we shall start by considering
\be
\tilde C(\tilde t) =e^{{-\tilde t\,}^p}\,,\label{expcutoff}
\ee
where $p$ is allowed to be an arbitrary positive real constant.
This also has the properties that it approaches 1 for small $\tilde t$, and
it goes rapidly to zero at large $\tilde t$.  Indeed, it clearly ensures
that the integration is convergent at large $t$.  With this exponential
choice for the cutoff, we have
\be
k \fft{d C_k(t)}{dk} = -p \omega \, t^p\, k^{p\omega}\, e^{-t^p k^{p\omega}}\,,
\label{kdkexp}
\ee
and so if we plug this and the asymptotic expansion for $Y(t)$, namely
\be
Y(t) \sim \sum_n Y_n\, t^n\,,\label{asymp}
\ee
into (\ref{betafunctional}), we get
\bea
\beta &=& \ft12 p\omega\, k^{p\omega}\,
\sum_n Y_n\, \int_0^\infty t^{p+n-1}\, e^{-t^p k^{p\omega}}\, dt\,,\nn\\
&=& \ft12\omega\, \sum_n k^{-n\omega}\, Y_n\, \int_0^\infty u^{n/p}\,
   e^{-u}\, du\,,\nn\\
&=& \ft12\omega\, \sum_n Y_n\, k^{-n\omega}\,
\Gamma\Big(\fft{n}{p}+1\Big)\,.\label{betaexp}
\eea

   Recalling that the asymptotic expansion (\ref{asymp}) for $Y(t)$ runs
over a discrete semi-infinite set of values for $n$, with $n\ge n_0$ where
$n_0$ is some negative number, we see that in order to get UV convergence of
all the integrals in (\ref{betaexp}), we must choose the constant $p$ in
the cutoff function (\ref{expcutoff}) such that
\be
p > |n_0|\,.
\ee
In our case, the most negative $n_0$ that we encounter in any of the
heat kernel expansions is $n_0=-3/2$, and so for our purposes it suffices
to take $p=2$.

It is interesting to compare the expansion for the beta function obtained
in the last line of (\ref{betaexp}) with the one for the step-function
cutoff, which follows from (\ref{genbeta}):
\be
\beta = \ft12\omega \sum_n Y_n\, k^{-n\omega}\,.\label{betatheta2}
\ee
Unsurprisingly, the terms with $n\ne0$ (which are scheme dependent) differ
when different cutoffs are chosen.  Note, however, that the $Y_0$ term
in (\ref{betatheta2}) is identical to the $Y_0$ term in (\ref{betaexp}),
for any non-zero choice of $p$.  One advantage of the theta-function cutoff
is that the $\beta$-function can be given, as in (\ref{genbeta}), as
a closed-form expression in terms of $Y(t)$.

\section{Euclideanization Rules}

For the details of the continuation of AdS$_3$ to $S^3$
and harmonic expansions on $S^3$, see \cite{Deger:1998nm}.

The AdS$_3$ metric is
\be
ds^2=-\cosh^2\rho dt^2+d\rho^2+\sinh^2\rho d\phi^2\ .
\ee
Our rule for Euclideanization is $\rho\mapsto i\rho$, which gives
\be
ds^2\mapsto -(\cos^2\rho dt^2+d\rho^2+\sin^2\rho d\phi^2)=-ds_{(E)}^2\ ,
\ee
which is locally the metric of the three-sphere with negative-definite signature.
The Ricci scalar of AdS$_3$ is equal to minus the Ricci scalar of
the standard positive-definite metric on the three-sphere.
Therefore, the rules for transforming equations on the AdS$_3$ background
to equations valid on the three-sphere background are
\be
g^{AdS}_{\mu\nu}\mapsto - g^{S^3}_{\mu\nu}\ ;\qquad
R^{AdS}\mapsto -R^{S^3}\ ;\qquad
\Lambda\mapsto -\Lambda\ .
\ee

The Dirac equation for a Majorana spinor on AdS$_3$ is $(\slashed D+m)\Psi=0$.
The Euclidean continuation of the (flat space) Dirac matrices is
\be
\gamma^0\mapsto i\gamma^0_{(E)}\ ,\qquad
\gamma^1\mapsto \gamma^1_{(E)}\ ,\qquad
\gamma^2\mapsto \gamma^2_{(E)}\ .
\ee
At the same time, for the dreibein components
\be
e_t^0\mapsto e_t^{(E)0}\ ,\qquad
e_\rho^1\mapsto ie_\rho^{(E)1}\ ,\qquad
e_\phi^2\mapsto ie_\phi^{(E)2}\ .
\ee
where $e^{(E)}$ are the dreibein for the standard Euclidean-signature metric on $S^3$.
These transformations together imply the rule $\slashed D\mapsto i \slashed D^{(E)}$. Because the metric we use on $S^3$ is positive definite, we can no longer have Majorana spinors. However, as usual, we use the Euclidean signature only to compute determinants in spacetime, without doubling the degrees of freedom.

\section{Some Heat Kernel Checks}

In this appendix all calculations are performed directly in the Euclidean signature.
Consider the contribution of a fermion field to the beta function.
It can be computed in either of two ways:
from the heat kernel of the Dirac operator, or from the heat kernel of its square
\be
\Delta=\slashed D^2=-\Box+\frac{R}{4}
\ee
The former has eigenvalues $\pm\rho(n+\ft32)$ and multiplicity $(n+1)(n+2)$,
the latter $\rho^2(n+\ft32)^2$ and multiplicity $2(n+1)(n+2)$, with $n=0,1,\ldots$ in both cases.
The heat kernels can be computed as spectral sums, along the lines of section 3.1.
From the spectral sums of the Dirac operator one finds
\be
Y_{\slashed D}(u)=\frac{V}{\pi^2u^3}\left(2+\ldots\right)
\ee
whereas from the spectral sum of the eigenvalues of the squared Dirac operator one gets
\be
Y_\Delta(t)=\frac{V}{(4\pi t)^{3/2}}\left(2+\ldots\right)
\ee
The two results agree if we make the identification $t=\pi^{1/3}u^2/4$.
\footnote{There are significant differences in the next term of the expansion,
and it has been argued in \cite{dona} that only the former procedure
is correct.}

Next we check that the correct way of summing the contributions of different spin components
to the beta functions is to sum the heat kernels of the respective operators,
with coefficient one for the highest order part,
{\it i.e.} the coefficients given in \eq{cs1} and \eq{cs2} do not play a role.
We check this in the case of pure bosonic gravity in the gauge $\alpha=1$, 
in which case the operator acting on metric fluctuations is equal to \cite{higher}
\be
\Delta_h=({\mathbf 1}-{\mathbf P})\left(-\Box+\frac{2}{3}R-2\Lambda\right)
-\frac{1}{2}{\mathbf P}\left(-\Box-\frac{1}{3}R-2\Lambda\right)\ ,
\ee
where $P^{\mu\nu\rho\sigma}=\ft13 g^{\mu\nu}g^{\rho\sigma}$ projects on the trace
and ${\mathbf 1}-{\mathbf P}$ on the tracefree part of $h_{\mu\nu}$.
The heat kernel of an operator of the form $-\Box+{\mathbf E}$
can be computed from the standard formula
\be
Y(t)=\frac{1}{(4\pi t)^{3/2}}\int d^3x\,\sqrt{g}\,
{\rm tr}\left[{\mathbf b_0}+{\mathbf b_2} t+{\mathbf b_4} t^2+\ldots\right]
\ee
with
\bea
{\mathbf b_0}&=&{\mathbf 1}\\
{\mathbf b_2}&=&\frac{R}{6}{\mathbf 1}-{\mathbf E}
\eea
From here one finds
\bea
Y_{trace}(t)&=&\frac{V}{(4\pi t)^{3/2}}
\left[1+\left(\frac{1}{2}R+2\Lambda\right)\,t+\ldots\right]\ ,
\\
Y_{tracefree}(t)&=&\frac{V}{(4\pi t)^{3/2}}
\left[5+\left(-\frac{15}{6}R+10\Lambda\right)\,t+\ldots\right]\ .
\eea
The result for $Y_{trace}$ agrees with $Y_{\Delta_{(h)}}(t)$,
evaluated in the gauge $\alpha=1$, while $Y_{tracefree}(t)$
agrees with the sum $Y_{\Delta_{(h^{TT})}}(t)+Y_{\Delta_{(\xi^T)}}(t)+Y_{\Delta_{(\sigma)}}(t)$.
Note in particular that when one adds up the heat kernels of the
differentially constrained fields $h^{TT}$, $\xi^T$ and $\sigma$
the terms with half-odd powers of $t$ cancel out.
(The trace and tracefree parts are defined by purely algebraic conditions.)


\section{Properties of $\Gamma_k$}


The computation of the $\beta$-functions require only the logarithmic derivative of $\Gamma_k$ with respect to
$k$. Nonetheless it is useful to examine the effect of the cut-off procedure described above in the computation of $\Gamma_k$ itself. While $\Gamma_k$ is a divergent, and thus ill-defined quantity, the following formulae make sense after taking their $k$-derivatives. With the theta-function cutoff, the representation \eq{hkrep} becomes
\begin{equation}
\Gamma_k=S-\frac{1}{2}\int_0^{1/(ak^\omega)} \frac{dt}{t}\,Y(t)\ .
\end{equation}
By adding and subtracting a constant for each mode, we can rewrite this as
\begin{equation}
\Gamma_k=S-\frac{1}{2}\left[
\int_0^{1/(ak^\omega)} \frac{dt}{t}\,\sum_n\left(1-e^{-t\lambda_n}\right)
-\int_0^{1/(ak^\omega)} \frac{dt}{t}\,\sum_n 1\right]\ .
\end{equation}
The sum in the second integral can be interpreted as $\zeta_\Delta(0)$,
where $\zeta_\Delta(s)\equiv \sum_n\lambda_n^{-s}$ is the zeta function of the operator $\Delta$,
and the first integral can be performed explicitly in terms of the exponential integral $\mbox{Ein}(x)\equiv \int_0^x(1-e^{-t})t^{-1}dt$:
\begin{equation}
\label{khrep2}
\Gamma_k=S-\frac{1}{2}\sum_n \mbox{Ein}(\lambda_n/(ak^\omega))-\gamma(k)\ ,
\end{equation}
where $\gamma(k)=\zeta(0)\int_0^{1/(ak^\omega)} t^{-1}dt$ and $\zeta(s)$ is the standard Riemann zeta function, $\zeta(s)=\sum_{n\ge 1}n^{-s}$.
For $k\to 0$, $\mbox{Ein}(\lambda_n/(ak^\omega))\to \log(\lambda_n/(ak^\omega))$,
so this reduces to the standard determinant formula with eigenvalues
measured in units of $ak^\omega$, modulo the irrelevant infinite constant $\gamma(0)$.
For $k>0$, however, $\Gamma_k$ is not given as the logarithm of a determinant any more,
but writing $\mbox{Ein}(\lambda_n/(ak^\omega))=\log(\lambda_n/(ak^\omega) F(\lambda_n/(ak^\omega)))$,
where $F$ tends to one when $k\to 0$, we can still interpret $\Gamma_k$ as the logarithm of a determinant, but now
of a modified wave operator $\widetilde\Delta$, where the eigenvalues are weighted by the function
$F(\lambda_n/(ak^\omega))$. Note that the term $\gamma(k)$ contributes an infinite constant to the beta function
which is cancelled by a contribution of opposite sign coming from the second term
in \eq{khrep2}.


\section{Quasi-supersymmetry of Gauge Fixing Conditions}


Th gauge fixing conditions \eq{G} and \eq{F} are motivated by the property that they eliminate the mixing terms between lower spin components of the fluctuation fields.  Here we study their behavior under the rigid supersymmetry transformations that leave the background invariant and act on the fluctuation fields as
\bea
\delta h_{\mu\nu} &=& {\bar\epsilon}\gamma_{(\mu} \psi_{\nu)}\ ,
\nn\w2
\delta\psi_\mu &=&-\ft14\left( \nabla_\rho h_{\sigma\mu}\gamma^{\rho\sigma}
+m h_{\mu\nu}\gamma^\nu\right)\epsilon\ ,
\label{bsusy}
\eea
where $\epsilon$ is understood to be a Killing spinor. Varying the bosonic gauge condition \eq{G} under these transformations gives
\be
\delta G_\mu= {\bar\epsilon}F_\mu\ ,
\ee
where $F_\mu= \bar\epsilon (\slashed{D}-\frac52 m)\phi_\mu +\cdots$, with ellipses denoting terms depending on $\chi,\psi$ and their derivatives. A gauge condition that preserves supersymmetry about the supersymmetric background would require that $\delta G_\mu$ be proportional to $\beps\gamma_\mu F$. This is not the case here due to the presence of the $\phi_\mu$ dependent terms, which are nonvanishing on-shell as well. Nonetheless, we find that
\be
\gamma^\mu F_\mu= -\frac{\alpha}{3} \left(\slashed{D}-\frac32 m\right)\psi
+\frac{10}{3} \left(\Box +\frac{R}{8}\right)\chi\ .
\ee
Comparing this result with the action of $\nk$ on the gauge condition $F$ which gives
\be
\nk F =\frac{1}{\alpha'} \left( \slashed{D} -\ft32 \rho \right)\psi+
\left(\Box +\frac{R}{8}\right)\chi\ ,
\ee
we see that on shell, for which $\rho=m$, we have the relation
\be
\gamma^\mu F_\mu = -\frac{10}{3} \alpha \nk F\ .
\ee
provided that we choose
\be
\alpha =-10\alpha'\ .
\ee
A similar phenomenon has been encountered in \cite{Kojima:1993hb}, where $3D$ supergravity was quantized around Minkowski spacetime.

\end{appendix}

\end{document}